\documentclass{aa}
\usepackage{amsmath}
\usepackage{graphicx}
\usepackage{txfonts}
\usepackage{lscape}
\usepackage{placeins}
\usepackage{afterpage}
\usepackage{stfloats}
\usepackage{booktabs}
\usepackage{xcolor}
\usepackage{placeins}
\usepackage[breaklinks=true,colorlinks=true,
            linkcolor=blue,citecolor=blue,urlcolor=black]{hyperref}

\begin{document}

\title{Beyond Pebble Isolation} 
\subtitle{Diverse Pathways to Giant Planet Formation Across Stellar and Orbital Scales}

\author{
    Lorenzo Peerani\inst{1}\thanks{\small \email{lorenzo.peerani@uzh.ch}}
    \and Sho Shibata\inst{2}
    \and Ravit Helled\inst{1}
}

\institute{
    Institut f\"ur Astrophysik, Universit\"at Z\"urich, Winterthurerstrasse 190, CH-8057 Z\"urich, Switzerland\\
    \and
    Department of Earth, Environmental and Planetary Sciences, Rice University, 6100 Main MS 126, Houston, TX 77005, USA
}

\date{Received Month DD, YYYY; accepted Month DD, YYYY}

  \abstract
   {Giant planet formation requires reaching crossover mass, i.e., when the mass of the gaseous envelope becomes equal to the mass of the core, within the disc's lifetime. The formation process depends critically on the orbital distance  and stellar mass.}
   {We simulate planet formation via pebble accretion up to crossover mass around stellar hosts with masses of 0.1--1.5\,$M_\odot$ considering a range of formation locations, with and without Type~I migration.} 
   {We use a modified version of \texttt{MESA} that couples pebble accretion, gas accretion, and disc evolution.}
   {We find that cold/warm Jupiters form, whereas in-situ formation fails at short orbital separations: viscous heating raises the isolation mass enough to assemble adequate cores, but the accompanying high disc temperature prevents cooling and suppresses gas giant formation. This supports migration-based explanations for the origin of hot Jupiters. At large orbital distances, crossover can be reached before pebble isolation mass. This is possible due to efficient envelope contraction in the cold, low-opacity outer disc. Inferred core masses at crossover range between about 1 and 20\,$M_\oplus$.}
   {Pebble accretion accommodates multiple formation pathways. Giant planets can also have very small cores. Overall, different formation conditions significantly influence planetary growth and can explain the diversity in compositions and internal structures observed in the exoplanet population.}


   \maketitle
   \nolinenumbers

\section{Introduction}
Planet formation theory aims to explain the rich diversity of  the observed exoplanet populations. Gas giant planets play a key role in this picture: they are the clearest outcome of efficient accretion of both heavy elements and nebular gas, and understanding their formation pathways informs the broader spectrum of planetary architectures, including terrestrial, sub-Neptunian, and Neptune-class planets. Whether a planetary system forms a gas giant, a super-Earth, or an intermediate-mass planet depends on the efficiency of the successive stages of accretion and the local conditions within the protoplanetary disc \citep{Ida_2004, Mordasini_2012, Bitsch_2015, Cridland_2018, Emsenhuber_2021}.

The leading framework for a bottom-up gas giant formation is core accretion \citep{Pollack_1996}: solid material (heavy elements) hierarchically assembles into a planetary core that subsequently accretes a gaseous (hydrogen-helium) envelope from the surrounding nebula. Within this framework, giant planet formation via pebble accretion has been suggested to allow a rapid growth of the core \citep{Weidenschilling_1977, Brauer_2008,Ormel_2010, Birnstiel_2011, Lambrechts_2012}. In this scenario, a heavy-element core grows from millimetre-to-centimetre-scale pebbles that  decouple from the gas and drift radially inward due to the sub-Keplerian headwind and (Phase 1). Core growth terminates at the pebble isolation mass when the planet opens a partial gap that reverses the local pressure gradient and halts inward pebble drift (Phase 2) \citep{Lambrechts_2014, Bitsch_2018}. Then, gas accretes quasi-statically on the Kelvin-Helmholtz timescale,  governed by envelope opacity, until the envelope mass approaches the core mass (i.e., crossover mass) and hydrostatic equilibrium breaks down, triggering runaway gas accretion until the local nebular reservoir is exhausted (Phase 3) \citep{Ikoma_2000, Lambrechts_2019, Ginzburg_2019}. Halting at any of these stages leads to the formation of a variety of planetary masses and compositions. 
\par
A fundamental constraint on theories of giant planet formation is that the entire formation process must be completed before the protoplanetary disc disperses, which typically occurs within a few million years (Myr) \citep{Mamajek_2009, Ribas_2015, Pfalzner_2026}. 
Although planet formation cannot be directly observed, current observations play a key role in constraining formation models. For example, ALMA characterises the dust and gas structure of protoplanetary discs, informing the physical conditions available for planet formation \citep{Keppler_2019, Andrews_2020, Close_2025, Lawlor_2026}.  JWST's atmospheric characterisation of exoplanets encodes present-day compositional signatures that can be linked back to formation history that can also inform theoretical frameworks \citep{Bitsch_2015,Chapman_2017, Booth_2019, Chachan_2021, Pacetti_2025}. The statistical diversity of the exoplanet population itself also serves as an important ensemble test for planet formation theory, despite the obvious observational biases. 
\par  
Population synthesis models based on planetesimal accretion have successfully
reproduced key statistical properties of the exoplanet population, including
giant planet occurrence rates and the period--mass distribution
\citep{Mordasini_2009, Burn_2024, Kimura_2026}, and have been extended to
low-mass stars \citep{Burn_2021}. Formation models that focus on planetary growth via pebble accretion 
have also been developed \citep{Liu_2020, Pan_2024}. However, the disc scaling relations
and pebble flux evolution around low-mass stars remain poorly constrained \citep{Long_2022}.

The occurrence rate of cold Jupiters is observed to depend on stellar mass \citep{Johnson_2010, Ghezzi_2018, Fulton_2021}, with the protoplanetary disc environment regulating both core growth and subsequent gas accretion \citep{Shibata_2025, Pan_2025}. Discs around M-dwarfs are thought to be smaller and  less massive \citep{Andrews_2013,Pascucci_2016, Manara_2022}, which reduces the total pebble flux available for core formation. The lower dust and gas masses of the  discs limit the material available for both core growth and subsequent gas accretion. Lower accretion rates lead to slower growth which in turn results in longer timescales to reach runaway gas accretion \citep{Papaloizou_1999}.

Giant planet formation around M dwarfs is therefore a race against disc dispersal under intrinsically unfavourable initial conditions, making it a critical test of any planet formation theory \citep{Johnson_2010,Liu_2020,Burn_2021}. 

Previous theoretical studies have identified the key parameters governing giant planet formation around low-mass stars \citep{Pan_2024, Sanchez_2025}. Observations have now confirmed the existence of gas giants orbiting M dwarfs \citep{Morales_2019, Bryant_2023, Kanodia_2024, Bryant_2025}, challenging standard core models and motivating a systematic exploration of the underlying formation physics. 

In this work, we investigate how planet formation via pebble accretion changes under different formation conditions, in particular, stellar hosts, formation location, and migration history. Our paper is organised as follows. In Section~\ref{sec:methods} we describe the planet formation simulations. In Section~\ref{sec:results} we present the results: the embryo formation timescales as a function of stellar mass and orbital distance, the crossover-mass timescales compared with the observed giant-planet population, and representative accretion tracks computed with and without migration. We discuss our results in Section~\ref{Discussion}, and present our conclusions in Section~\ref{sec:conclusion}.  

\section{Methods}
\label{sec:methods}

\subsection{Planet formation model}

To simulate planetary formation, we used \texttt{mespa}, an extension of \texttt{MESA} (Modules for Experiments in Stellar Astrophysics) that was modified for planetary modelling \citep{Helled_2025}. This allows to model the  planetary evolution and formation processes \citep[e.g.,][]{Valletta_2020,MolLous_2024}. We adopt the planet accretion framework developed by \citet{Shibata_2025}, which self-consistently couples pebble accretion, gas accretion, and disc evolution across different stellar environments. 
The description of the model can be summarised as follows:
\begin{itemize}
    \item \textit{Disc evolution:} modelled by a combination of stellar irradiation and viscous heating. At small separations, the midplane temperature is viscously heated, while at larger distances stellar irradiation dominates (Fig.~\ref{fig.isolation_masses}). The gas surface density declines over time as the disc undergoes viscous accretion. Consequently, the midplane temperature decreases alongside the surface density, which governs the viscous heating rate. The disc's properties scale with stellar mass. The full expressions are provided in Appendix~\ref{appendix:disc_model}.
    \item \textit{Pebble Accretion:} pebble accretion is modelled following the framework of  \citet{Lambrechts_2014, Johansen_2017}. Further descriptions of the physics can be found in \citet{Shibata_2025} or see Section~\ref{sec:accretion} in the Appendix.
    \item \textit{Gas Accretion:} 
    Gas accretion is set by the planet's accretion radius, which arises from a competition between the Bondi (thermal) radius and the Hill (tidal) radius. The two are combined harmonically, the accretion radius is explained as $R_{\rm acc}^{-1} = R_B^{-1} + 4\,R_H^{-1}$ \citep{Lissauer_2009}, so that $R_{\rm acc}$ approaches whichever term dominates: for low-mass planets $R_B \ll R_H/4$ and accretion is thermally (Bondi) limited, whereas for massive planets $R_H/4 \ll R_B$ and it becomes tidally (Hill) limited. 
    \item  \textit{Isolation Mass:} perturbation of the gas disc due to the planet's size is included. The pebble isolation mass can be expressed as \citep{Bitsch_2018}:
\begin{align}
M_{\rm iso} =\;& 25.0\,\mathrm{M}_\oplus
\left( \frac{M_\star}{\mathrm{M}_\odot} \right)
\left( \frac{H_{\rm gas}/r}{0.05} \right)^3 \nonumber \\
&\times
\left(
0.34 \left[\frac{-3}{\log_{10}(\alpha_{\rm turb})}\right]^4 + 0.66
\right)
\left(
1 - \frac{\displaystyle \frac{\partial \ln P_{\rm disc}}{\partial \ln r} + 2.5}{6}
\right),
\label{eq.isolation}
\end{align}
where $h = H_{\rm gas}/r$ is the disc aspect ratio.
\begin{figure}[!h]
    \centering
    \includegraphics[width=0.5\textwidth]{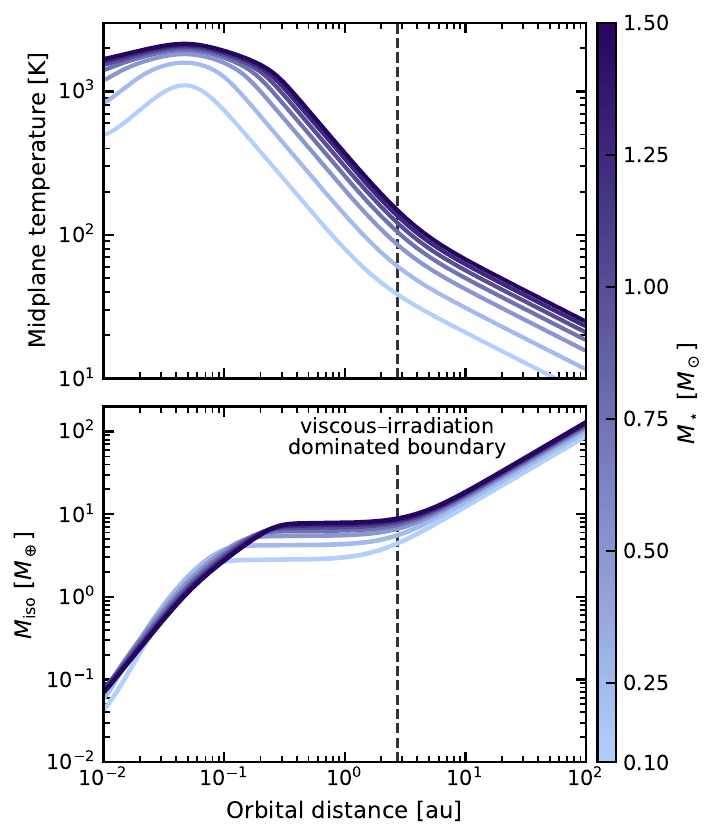} 
    \caption{Disc thermal profiles at the midplane for different stellar masses for an irradiated+viscous disc. Below, pebble isolation masses as a function of  orbital distance. The further away from the star, the larger the isolation mass is. The shading corresponds to the different stellar hosts' masses. The dashed line marks the switch between viscous-dominated heat to irradiated, located at around 2.7 AU (1.6 AU or 3.4 AU respectively for a 0.1 and 1.5 $\mathrm{M}_{\odot}$ stellar host). The values have been taken at time $t$ = 1 Myr as an illustrative snapshot (both the midplane temperature and the isolation mass), the \citet{Shakura_1973} viscosity parameter} $\alpha$ = $10^{-3}$ which controls the disc's evolution, and $\alpha_{turb}$ = $10^{-4}$ which determines the gas turbulence and vertical distribution of pebbles.
    \label{fig.isolation_masses}
\end{figure}
    Fig.~\ref{fig.isolation_masses} shows how the isolation mass $M_{\rm iso}$ changes with stellar mass and orbital distance. From  Eq.~\ref{eq.isolation} it is clear that the isolation mass scales steeply with the disc aspect ratio:  $M_{\rm iso}\propto (h/r)^3$. In the outer disc the aspect ratio is set by stellar irradiation, while at small separations viscous heating dominates and inflates the disc, keeping the isolation mass roughly constant towards the inner edge. 
    Pebble inhibition when hitting isolation mass is modelled with an exponential formulation as follows:

\begin{equation}
    \dot{M}_\mathrm{Z} = \min\!\left( \dot{M}_\mathrm{peb},\; 
    \dot{M}_\mathrm{peb,acc}\, 
    \exp\!\left( -\left[ \frac{M_\mathrm{p}}{M_\mathrm{iso}} \right]^{10} 
    \right) \right),
    \label{eq:mz_dot}
\end{equation}
where $\dot{M}_{\rm peb}$ is the inward pebble mass flux supplied by the
disc at the planet's location, $\dot{M}_{\rm peb,acc}$ is the rate at which the
planet gravitationally captures pebbles from that flux, and $M_{\rm p}$ is the
planet's mass. Taking the minimum ensures that the planet never accretes solids faster
than they drift past its orbit. The exponential factor suppresses accretion as
the planet approaches $M_{\rm iso}$ rather than truncating it: filtering at the
pressure bump is size-dependent, so small, well-coupled grains continue to reach
the planet, and a smooth cut-off avoids the discontinuity in the accretion energy
released at the core--envelope boundary.
    \item \textit{Type I Migration:} Lindblad and corotation torques have been modelled with the prescription from \citet{Paardekooper_2011}. The full torque expressions and migration formalism are provided in Appendix~\ref{appendix:migration}.
\end{itemize}

\subsection{Initial Embryo Formation Timescale}
\label{sec:formation_timescales}

The formation simulations with \texttt{mespa} are initialized with a seed mass of $0.1\,\mathrm{M}_\oplus$, corresponding to a planetary embryo that has already undergone significant solid growth. To physically motivate the formation timescale of such an embryo, we estimate the initial planetesimal mass produced by the streaming instability and subsequent pebble accretion.

We follow the framework of \citet{Lau_2022}, in which the characteristic maximum planetesimal mass formed via the streaming instability is given by the following formula taken from \citet{Klahr_2020}:
\begin{align}
M_{\rm seed} \sim 7.22 \times 10^{-3}
\left( \frac{\delta / \tau_{\rm f}}{10^{-2}} \right)^{3/2}
\left( \frac{H_{\rm gas}/r}{0.058} \right)^3
\mathrm{M}_\oplus, 
\end{align}
where $\delta$ is the turbulent diffusivity of the gas, $\tau_{\rm f}$ is the Stokes number of the particles participating in the streaming instability, $H_{\rm gas}$ is the scale height, and $h = H_{\rm gas}/r$ is the disc's aspect ratio. We approximated $\tau_{\rm f}$ to 0.01 and set $\delta$ to $10^{-4}$. This value of $\tau_{\rm f}$ enters only in the seed-mass estimate; during the simulations the Stokes number varies and is computed at each timestep from the
local disc conditions (Appendix~\ref{sec:accretion}). The scale height is computed self-consistently from the disc's temperature structure.

From this initial planetesimal mass, which is significantly smaller than the  $0.1\,\mathrm{M}_\oplus$ we used for the first  simulation, we estimate the time required to grow to $0.1\,\mathrm{M}_\oplus$ via pebble accretion. This calculation is performed outside \texttt{mespa} and sets the epoch $t_0$ at
which each simulation begins (and not the initial mass). During this early stage, gas accretion is negligible, and the planetary growth is dominated by the accretion of inward-drifting pebbles. The accretion rate is determined by the local pebble flux and accretion efficiency, following the prescription described in Sect.~\ref{sec:accretion}. We assume that gas giant formation proceeds from the estimated maximum planetesimal mass. Although this assumption may  underestimate the protoplanetary embryo growth timescale, it is physically motivated by the expectation that gas giants emerge from early embryos, those that undergo exceptionally efficient accretion \citep{Pollack_1996}.

This approach allows us to map the formation timescale of planetary embryos as a function of stellar mass and orbital distance. In particular, lower pebble fluxes at large orbital distances and around low-mass stars lead to significantly longer growth timescales, which can delay the onset of envelope accretion and influence whether planets reach the pebble isolation mass \citep{Ormel_Vazan_2021}. 

\subsection{Simulations Setup}
We performed simulations for a wide range of stellar masses and initial orbital distances. To define a consistent scaling of orbital distance with stellar mass, we adopted the relation of
$a_{\rm step} = 2.67 \left( M_\star / \mathrm{M}_\odot \right)^{1/2}~\mathrm{AU}$. Such spacing effectively allows smaller steps that are needed for the restrictive  ranges of planet formation around smaller stars. For smaller separations, we tested cases with orbital distances of 0.01 and 0.05 AU.

Our simulations were carried out across stellar masses of 0.1, 0.25, 0.5, 0.75, 1.0, 1.25, and 1.5 $\mathrm{M}_{\odot}$, sampling in a wide orbital range. Simulations were performed until the crossover mass was  reached (i.e., when $M_{\rm Z}$ = $M_{\rm H-He}$). If this is the case, we can assume that a gas giant can form. We started the simulations at $t_{0}$, where $t_{0}$ is informed by physical timescales to reach 0.1 $\mathrm{M}_\oplus$, as explained in Section~\ref{sec:formation_timescales}. 
The simulations begin with a heavy-element core of 0.1 $\mathrm{M}_{\oplus}$. We assumed that the infalling pebbles are added to the core (i.e., no envelope pollution). The resulting gravitational heat is then  released at the core-envelope boundary. This is clearly a simplification, as atmospheric pollution with heavy elements can affect the planetary growth in various ways  \citep{Hori_2010, Venturini_2015, Valletta_2020, MolLous_2024}. For the atmosphere, we assumed a low grain opacity factor of $f_{g}$ = 0.01. A low grain opacity value is expected from grain growth and settling \citep{Pollack_1996, Movshovitz_2010, Mordasini_2014}. We assumed a Solar-like disc metallicity of 0.02, and a disc viscosity $\alpha$ of $10^{-3}$, followed by \citet{Sanchez_2025}, and motivated by observational and theoretical constraints \citep{Teague_2016, Simon_2018, Rosotti_2023}. This standard $\alpha$ viscosity
drives the disc's evolution and viscous heating, and is different  from the turbulent diffusion parameter $\alpha_{\rm turb} = 10^{-4}$ which sets the vertical stirring of pebbles (Appendix~\ref{sec:accretion}). We performed simulations with and without Type I migration to better understand the inferred trends and compare these two formation scenarios. The parameters used in this study and a discussion on their expected effect on the results are presented in Appendix~\ref{appendix:assumptions}. 

\subsection{Disc lifetime}
\label{disc_lifetime}
The simulations were evolved over several million years, after which only physically 
meaningful outcomes are retained through a filtering criterion based on the disc dispersal timescale. Since the disc's lifetime strongly influences whether a protoplanet can reach crossover mass, a physically motivated upper limit must be imposed. Disc dispersal timescales vary considerably across stellar systems \citep{Gorti_2006, Hillenbrand_2008}. \citet{Kennedy_2009} established an inverse relationship with stellar mass, and \citet{Pfalzner_2026} showed that the intrinsic scatter is too large to be captured by a single power law. 
Nonetheless, we constructed a piecewise scaling relation taken from the mode (peak) of the disc's lifetime distributions of \citet{Pfalzner_2026}. 
These distributions are given in three broad stellar-mass bins. We collapse each bin to a single representative mass and take its mode as the characteristic disc lifetime: the low-mass bin ($0.01$--$0.2\,\mathrm{M_\odot}$, $7.20\,\mathrm{Myr}$) at 
$0.1\,\mathrm{M_\odot}$, the intermediate bin 
($0.2$--$1.0\,\mathrm{M_\odot}$, $7.66\,\mathrm{Myr}$) at $1.0\,\mathrm{M_\odot}$, 
and the high-mass bin ($1.0$--$3.0\,\mathrm{M_\odot}$, $3.72\,\mathrm{Myr}$) 
also at $1.0\,\mathrm{M_\odot}$. These three points anchor a two-branch power law split at the physical X-ray/FUV transition ($\sim 1\,\mathrm{M_\odot}$); a single transition is sufficient here, since we only consider  
$0.1$--$1.5\,\mathrm{M_\odot}$. Below $1\,\mathrm{M_\odot}$ the slope is fixed 
by the two lower bins, giving a shallow index of $+0.03$: the disc's  lifetime is essentially mass-independent in this X-ray-photoevaporation--dominated regime \citep{Picogna_2021}. Above $1\,\mathrm{M_\odot}$ only a single mode is 
available, setting the normalisation ($3.7\,\mathrm{Myr}$ at $1\,\mathrm{M_\odot}$).  The slope over our $1.0$--$1.5\,\mathrm{M_\odot}$ range is adopted from the inverse mass--lifetime trend $\tau_\mathrm{disc} \propto M_*^{-0.5}$ of \citet{Kennedy_2009}, reflecting the stronger FUV-driven winds around higher-mass stars \citep{Komaki_2021}. The disc lifetime distributions we used  are: 

\begin{equation}
    \tau_\mathrm{disc}(M_{\star}) = 
    \begin{cases}
        7.2\,\mathrm{Myr} \times 
        \left(\dfrac{M_{\star}}{0.1\,\mathrm{M_\odot}}\right)^{+0.03} 
        & M_{\star} < 1\,\mathrm{M_\odot} \\[10pt]
        3.7\,\mathrm{Myr} \times 
        \left(\dfrac{M_{\star}}{1.0\,\mathrm{M_\odot}}\right)^{-0.5} 
        & M_{\star} \geq 1\,\mathrm{M_\odot},
    \end{cases}
    \label{eq:disc_lifetime}
\end{equation} 
suggesting disc lifetimes of 7.2, 3.7, and 3.02 Myr for stars with masses of  0.1, 1, and 1.5 $\mathrm{M}_\odot$, respectively. 

We deliberately chose the mode rather than the median to adopt the most conservative possible lower value on disc lifetime and avoid inflating our results. Since the underlying disc lifetime distributions are right-skewed, the mode lies below the median and represents the most 
probable dispersal time rather than the average. This choice 
preferentially rejects long-lived outliers, minimising the risk of 
retaining runs whose outcomes are driven by unrealistically extended disc lifetimes. Because $\tau_\mathrm{disc}$ enters our analysis only as a ``truncation limit" and not as a driver of the planet evolution, our results are insensitive to the exact values of these exponents.

\section{Results}
\label{sec:results}

\subsection{Embryo formation timescales} 
\begin{figure}[!ht]
    \centering
    \includegraphics[width=0.45\textwidth]{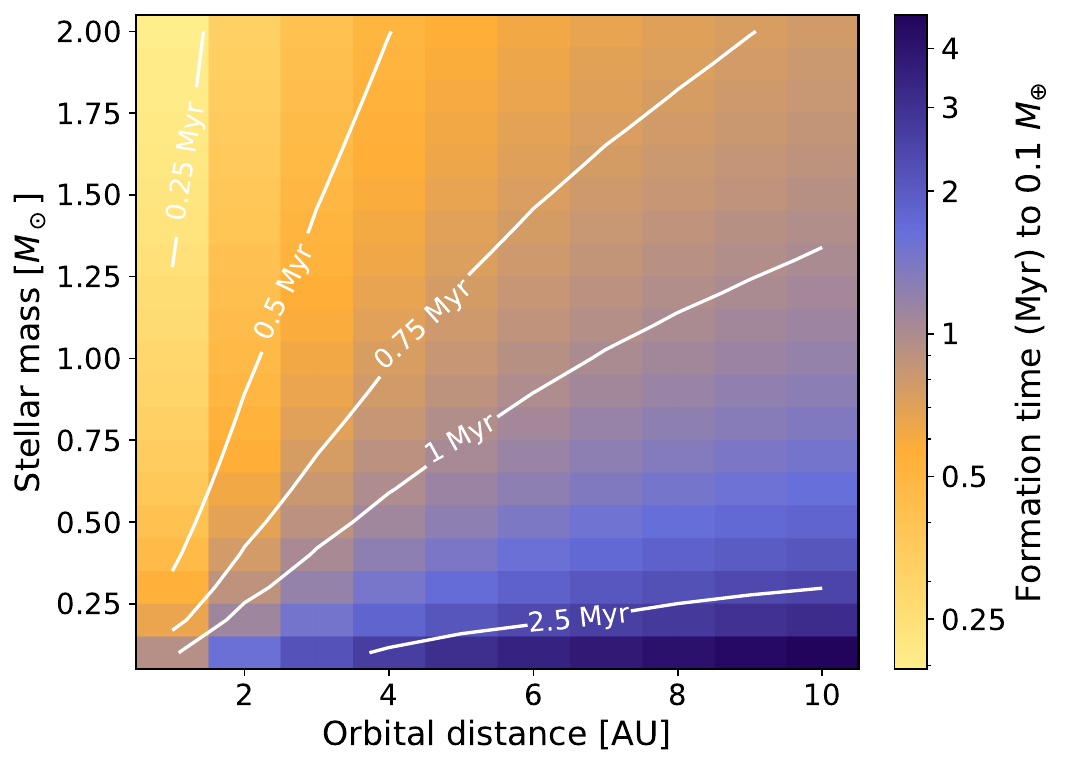} 
    \caption{Formation timescale (in log-scale) for planetary embryos at 0.1 $\mathrm{M}_{\oplus}$ as a function of stellar mass ($\mathrm{M}_{\odot}$) and orbital distance (AU). The white curves show interpolated constant formation times at 0.25, 0.5, 1, and 2.5 Myr. } 
    \label{fig.initial_timescale}
\end{figure}

The results for the planetary growth using  the approach described in Section~\ref{sec:formation_timescales} are shown in Fig.~\ref{fig.initial_timescale}. 
The onset of planet formation is determined  by the streaming instability and strongly depends on both orbital distance and stellar mass. In particular, formation timescales increase significantly with increasing radial distance from the host star. This effect becomes more pronounced for lower-mass stars. Therefore, the formation timescales already highlight  the first challenge associated with forming gas giants around low-mass stars. For example, at $\sim$10 AU around a $0.1\,\mathrm{M}_{\odot}$ star, the time required to form a $0.1\,\mathrm{M}_{\oplus}$ embryo already exceeds $\sim$5 Myr. This optimal region scales with a steep power law with stellar mass. 

These initial timescales represent a critical bottleneck: if the formation of the planetary embryo is delayed, the available time for subsequent growth (set by the disc's lifetime) is severely reduced, which directly impacts the likelihood of reaching crossover mass and forming a gas giant planet. 

\subsection{Crossover-mass timescales and the observed population}
\begin{figure*}[!ht]
    \centering
    \includegraphics[width=1\textwidth]{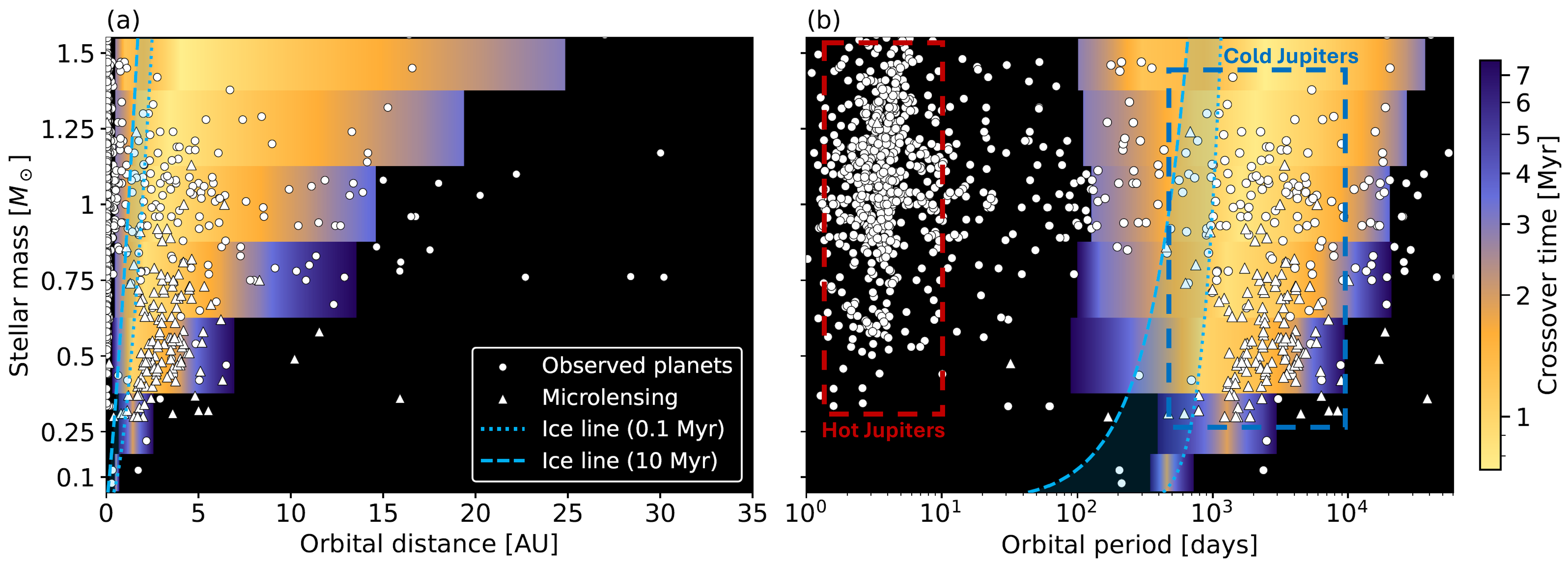} 
    \caption{Colourmaps showing the predicted timescales  to reach crossover mass assuming in-situ formation as a function  of formation location and stellar mass ($\mathrm{M}_{\odot}$). Panel (a) is plotted over orbital distance (AU) and panel (b) in log orbital period (days). The black background corresponds to timescales longer than the assumed disc lifetime ($\tau_{\rm disc}$ as discussed in Section~\ref{disc_lifetime}). The white dots show the observed gas giants above 0.25 $\mathrm{M}_{\rm J}$ from various surveys. The triangles are planets detected by microlensing.  Observational data from  the \texttt{exoplanet.eu} database. The light blue dotted lines indicate the water iceline of the disc model used here (Appendix~\ref{appendix:disc_model}), shown at 0.1 and 10~Myr.} In panel (b) we also indicate the two main regions hot and cold Jupiters. 
    \label{fig.colourmap}
\end{figure*}

Figure~\ref{fig.colourmap} shows the theoretical crossover mass formation timescales as a function of stellar mass and orbital distance/period, overlaid with confirmed gas giants from the \texttt{exoplanet.eu} catalogue\footnote{\url{http://exoplanet.eu}}, comprising radial velocity, microlensing, direct imaging, and transit detections. In these simulations, the orbital distance is kept fixed to isolate its impact on the formation efficiency. 

The resulting map reveals a clear trend: for a given  stellar mass, there is an optimal range of orbital distances where gas giant formation is most efficient. Around low-mass stars, this favourable region is narrower and shifted towards smaller orbital distances, while the corresponding formation timescales are systematically longer. Planets forming around low-mass stars reach crossover later, a trend that is primarily controlled by the extended timescales to reach $0.1\,\mathrm{M}_\oplus$ shown in Fig.~\ref{fig.initial_timescale}. This is supported by the smaller number of gas giants observed around low-mass stars (shown as fewer point dots in Fig.~\ref{fig.colourmap}).

The range of solutions shown in the colourmap reflects the balance between 
competing physical processes. At large orbital distances, longer growth timescales arise from the longer orbital timescales, the reduced gas surface densities, and slower pebble growth and drift (i.e.\ smaller Stokes numbers). At small 
orbital distances, the planetary growth is limited by the reduced solid content interior to the snowline, where pebbles lose their ices. This limitation is not absolute: at a fixed metallicity and a fixed disc-mass-to-star-mass ratio (2\% and 10\% respectively), more massive stars have a greater flux of solids, allowing giant planets to reach crossover mass even interior to the ice line, as shown in Fig.~\ref{fig.colourmap}. The interplay between these effects leads to a 
``sweet spot'' where planet formation is most efficient. 

From the 1003 gas giants above $0.25\,\mathrm{M}_{\rm J}$ plotted, only 23 (2.3\% of the filtered data) are located beyond the predicted formation boundary. These outliers may reflect disc lifetimes longer than the median scaling adopted here, or variations in the disc parameters that, for simplicity, were held fixed in this study. It is also possible that the objects at large  orbital distances (beyond 30-50 AU) were formed by disk instability  \citep{Baehr_2023, Helled_2026}.  
The apparent excess of giant planets around lower-mass stars is dominated by microlensing 
detections (triangles in Fig.~\ref{fig.colourmap}, \citealt{Gaudi_2012, Suzuki_2018}). These surveys, however, 
sample the M-dwarf-dominated field population rather than the bright FGK hosts targeted by radial-velocity and transit surveys. Therefore, this excess is affected by observational biases and probably reflects the host samples probed by 
each technique rather than a higher formation efficiency around low-mass stars, and the host masses themselves may be underestimated \citep[for example,][]{Nunota_2024, Bennett_2024}. 
The apparent depletion of observed giants around higher-mass stars is likely an observational bias rather than a physical suppression: our models predict that the formation window widens around $1.5\,\mathrm{M}_{\odot}$ hosts due to their more massive discs providing a broader range of orbital distances where crossover can be reached within the disc's  lifetime. This result is  consistent with the occurrence trend reported by \citet{Johnson_2010}. We note  that massive stars ($\gtrsim 1.3\,\mathrm{M}_\odot$) tend to rotate fast, which broadens their spectral lines making radial-velocity detection more challenging \citep{Galland_2005, Johnson_2008}. This effect is expected to bias the observed occurrence rates against high-mass hosts.
Panel (b) of Figure~\ref{fig.colourmap} illustrates the limitation of in-situ formation. Although the cold Jupiter formation region is well reproduced, the model cannot account for the orbital evolution required to explain hot Jupiters. The absence of viable in-situ formation solutions at small separations is
informative, as it supports the inefficiency of in-situ hot Jupiter formation: 66.1\% of the plotted gas giant population lies closer to the star than the in-situ model can explain. This points to an inward transport mechanism, either disc-driven migration or high-eccentricity migration following planet--planet
scattering \citep{Rasio_1996, Chatterjee_2008, Dawson_2018}. We note that our simulations follow single planets and therefore cannot simulate the latter mechanism.

\subsection{In-situ formation}
\label{Migration}
The accretion histories reveal the strong dependence of the formation timescales and crossover mass on both stellar mass and orbital distance.  Figure~\ref{fig.masses} compares various growth tracks for simulations within the disc dissipation timescale constraint. The cases assuming in-situ formation are shown in the four upper panels. We show  the planetary masses vs.~orbital distance for different stellar hosts. The evolving planetary metallicity $Z$ corresponds to the core-to-envelope ratio. For the majority of the in-situ formation cases (yellow-to-blue colour palette), for a given stellar mass, increasing the orbital distance initially drives higher crossover masses through isolation mass physics: as the isolation mass increases (the dashed line in Fig.~\ref{fig.masses}), and since solid accretion is suspended once isolation is reached, the crossover mass tracks as $M_\mathrm{crossover} \gtrsim 2M_\mathrm{iso}$.

This trend does not persist indefinitely. Fig.~\ref{fig.masses} shows that crossover mass reaches a maximum at intermediate separations and then declines at larger distances. This behaviour is evident across all stellar masses. Notably, at large orbital separations the simulations never reach the (dashed) isolation mass, instead attaining crossover before reaching pebble isolation mass. 
This turnover marks a change of regime: at large separations the pebble supply
is exhausted before isolation mass is reached, allowing the envelope to cool and
contract at a lower core mass. We further discuss this pebble-exhaustion regime in
Sect.~\ref{Outer_disc}.

\subsection{Migration}
\begin{figure*}[!tb]
    \centering
    \includegraphics[width=1\textwidth]{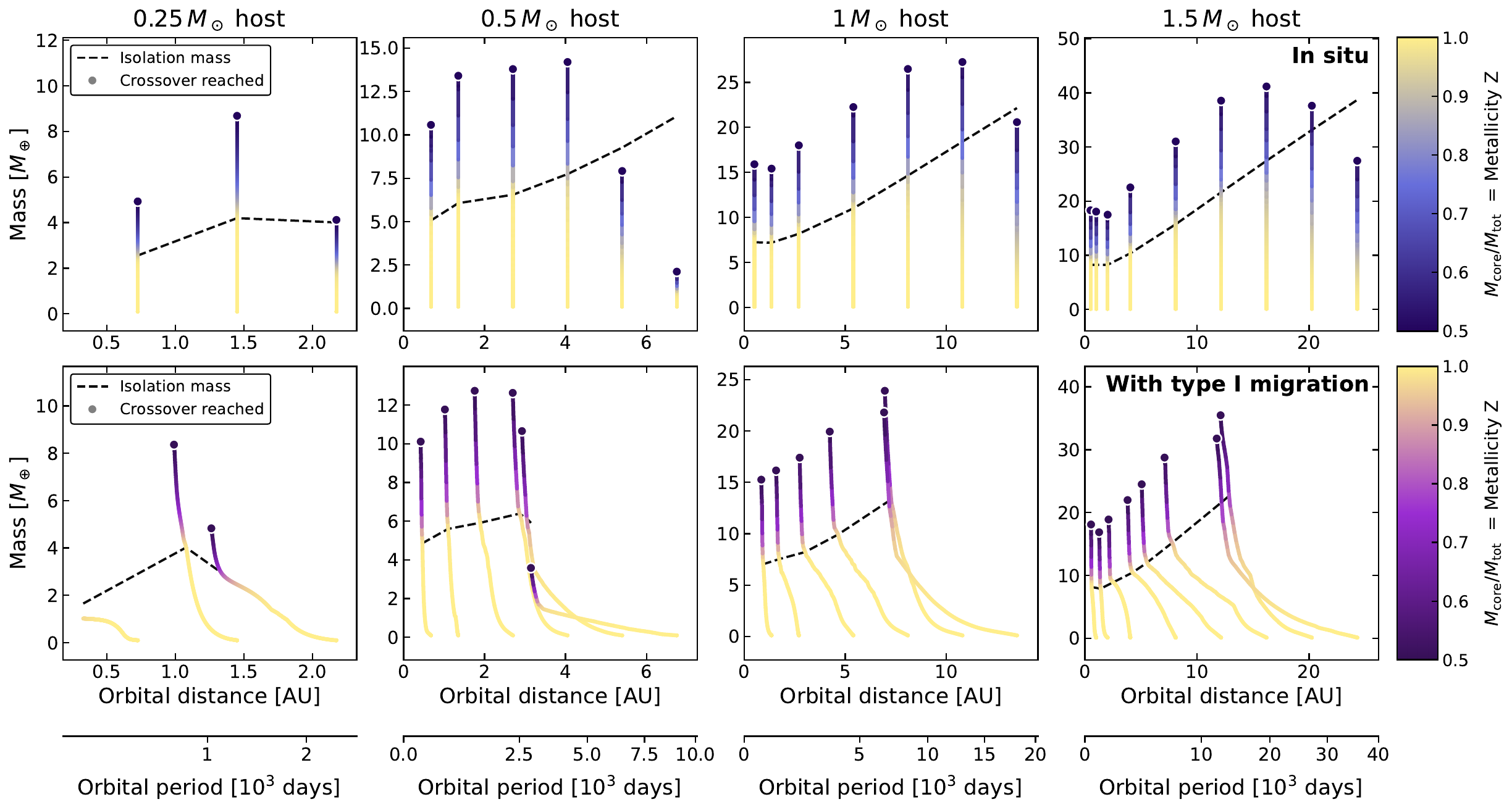} 
    \caption{Accretion tracks for host star masses of 0.25, 0.5, 1, and 1.5 $\mathrm{M}_{\odot}$, showing the total planetary mass ($\mathrm{M}_{\oplus}$) over orbital distance (AU) and orbital period (10$^{3}$ days)}. The evolving composition during accretion is shown as a colourmap, denoting the change in metallicity $Z$ as a ratio between the mass of the core and the mass of the total planet. The top row corresponds for the case of in-situ formation (yellow-to-blue colour scale) while the bottom one to formation followed by Type I migration (yellow-to-purple colour palette). The simulation is terminated when the planet reaches crossover mass. A case where crossover is reached is indicated with a black dot at the end of the track. The theoretical isolation mass according to eq.~\ref{eq.isolation} is plotted in a dashed black line. Note that the axes range are different for the different panels.
    \label{fig.masses}
\end{figure*}

Migration reduces the final crossover masses by shifting planets to smaller orbital radii where isolation masses are lower. The ``successful" formation locations are the same for migrating and in-situ embryos. The cases assuming migration are shown in the four bottom panels.  As Figure~\ref{fig.masses} shows, simulations that started from the same orbital distance reach crossover with comparable timescales regardless of whether Type~I migration is considered. Due to Type I migration, planets  can reach crossover mass since it is smaller at smaller orbital distances. If we had only considered migrating planets, the results presented in Figure~\ref{fig.colourmap} would have changed significantly. This is because no planets would have been left at large orbital periods and the cold population would have been significantly reduced. 
To estimate how far these planets would migrate once they open a gap we applied the model-independent growth track of
\citet{Tanaka_2020}. Because both the gas accretion rate and the Type~II migration speed scale with the gap surface density $\Sigma_{\rm gap}$, their ratio
$\mathrm{d}M_{\rm p}/\mathrm{d}\ln r$ is independent of the disc model, yielding an analytic evolution track in the mass--orbit plane (their Eq.~11):
\begin{equation}
\frac{a}{a_0} = \exp\!\left\{-\frac{3}{2}\left[\left(\frac{M_{\rm p}}{M_{\rm th}}\right)^{2/3}
- \left(\frac{M_0}{M_{\rm th}}\right)^{2/3}\right]\right\},
\label{eq.tanaka}
\end{equation}
where $(M_0,\,a_0)$ are the planet mass and orbital distance at the start of the track and
$M_{\rm th}=M_\star\,(0.29/6.0)^{3/2}\sim0.011\,M_\star$ is a threshold mass
($\sim 11\,M_{\rm J}$ at $1\,M_\odot$, scaling linearly with stellar mass). Physically, $M_{\rm th}$ corresponds to the mass scale where a planet begins to open a gap and migrate efficiently: planets below this mass undergo negligible Type II migration, while a planet that grows to $M_{\rm th}$ ends up at only ${\sim}1/5$ of its initial radius. 
For each stellar host, we take the innermost successful migrating protoplanet of Fig.~\ref{fig.masses}, adopting its crossover mass and orbital radius as $(M_0,\,r_0)$, and integrate Eq.~\ref{eq.tanaka} forward to final masses of $1$, $2$, and $10\,M_{\rm J}$, converting the resulting radii to orbital periods through Kepler's third law. Since these protoplanets are still far below $M_{\rm th}$ at crossover mass, the $(M_0/M_{\rm th})^{2/3}$ term is negligible and the planet experiences nearly all of its orbital displacement during its final, massive growth phase. The results are summarised in Table~\ref{tab.typeII}.  
Note that Eq.~\ref{eq.tanaka} gives a trajectory rather than a rate: the viscosity, which sets both the migration speed and the gas accretion rate
\citep{Savvidou_2023}, cancels from $dM_{\rm p}/d\ln r$.

Even under this optimistic, purely inward Type~II prescription, the migrated periods remain
long: a planet grown to $1\,\mathrm{M}_{\rm J}$ ends at $P\gtrsim70$~days across all stellar hosts, and only if
it reaches ${\sim}10\,\mathrm{M}_{\rm J}$ the period decreases to a few days for the 0.5 $M_{\odot}$ stellar host case which is inside the approximated $P\lesssim10$~days hot-Jupiter regime.  Around the other stellar hosts, the planetary periods fall to a few tens of days (i.e., outside the hot Jupiters region). We can therefore conclude that Type~II migration cannot fully explain the
observed close-in giants, in agreement with \citet{Tanaka_2020}, and reinforces the need for
an additional mechanism such as high-eccentricity migration to populate the shortest-period
orbits with planets. 
We note that other types of disk-driven migration, such as wind-driven migration \citep{Lega_2021} and Type III migration \citep[see e.g.][]{Nelson_2018} could also lead to the formation of hot/warm-Jupiters. However, this is expected to be  a minor formation pathways since hot-Jupiters are found to be rare.

\begin{table}
\centering
\caption{Orbital periods of the closest successful migrating protoplanet per
stellar host, before and after Type~II migration.}
\label{tab.typeII}
\renewcommand{\arraystretch}{1.2}
\begin{tabular}{ccccccc}
\hline\hline
$M_\star$ & $M_{\rm th}$ & $M_0$ & $P_0$ & $P_{1M_J}$ & $P_{2M_J}$ & $P_{10M_J}$ \\
{}[$\mathrm{M}_\odot$] & [$\mathrm{M}_{\rm J}$] & [$\mathrm{M}_{\rm J}$] & [d] & [d] & [d] & [d] \\
\hline
0.50 & 5.6 & 0.032 & 141 & 74 & 49 & 5 \\
0.75 & 8.3 & 0.046 & 654 & 406 & 295 & 55 \\
1.00 & 11.1 & 0.048 & 282 & 191 & 146 & 37 \\
1.25 & 13.9 & 0.055 & 824 & 591 & 471 & 143 \\
1.50 & 16.7 & 0.057 & 120 & 90 & 73 & 26 \\
\hline
\end{tabular}
\tablefoot{Protoplanets are those shown in Fig.~\ref{fig.masses}. $P_0$ is the
orbital period at crossover; $P_{1M_J}$, $P_{2M_J}$ and $P_{10M_J}$ are the
periods after Type~II migration to 1, 2 and 10~$\mathrm{M}_{\rm J}$, computed
with Eq.~11 of \citet{Tanaka_2020}. The initial masses $M_0$ are the crossover
masses obtained with \texttt{mespa}. The 0.1~$\mathrm{M}_\odot$ host is omitted
because no migrating run reached crossover mass within the disc lifetime.}
\end{table}

Compared to in-situ formation, formation with migration reach crossover mass below the isolation mass more rarely. This is because migration moves planets inwards, where the isolation mass is smaller and therefore more easily attained. The $0.5\,\mathrm{M}_{\rm \odot}$ case is an exception, still showing an anomalously low-mass core. The same effect would also occur around more massive hosts, provided the disc lifetime is sufficient to accommodate the slower formation timescales characteristic of larger orbital distances.

Among the migrating planets shown in Fig.~\ref{fig.masses}, one case behaves differently: around the 0.25~$\mathrm{M}_{\odot}$ host at 0.725~AU, the planet does not reach the pebble isolation mass, while its in-situ counterpart forms a giant planet. The in-situ planet stays at 0.725~AU, where the disc soon cools below the water ice line temperature, so it accretes icy pebbles efficiently and reaches crossover. The migrating planet instead moves inward faster than the ice line and stays in ${\sim}210$~K gas, where dry pebbles make solid accretion is less efficient. Its core stalls at ${\simeq}\,1\,\mathrm{M}_\oplus$, below the local isolation mass and too small to contract its envelope within the disc's lifetime.
As the planet accretes material, the accretion rate becomes Hill limited. Type I migration towards the star makes the Hill radius shrink, and part of the envelope becomes unbound. This leads to atmospheric loss of ${\sim}50$--$90\%$ for the 0.1 and 0.25~$\mathrm{M}_{\odot}$ migrating cases, by the time of disc dissipation. This effect on the Hill radius is stronger around low-mass stars, and therefore further hinders gas giant formation around low-mass stars.

\begin{figure*}[!ht]
    \centering
    \includegraphics[width=1\textwidth]{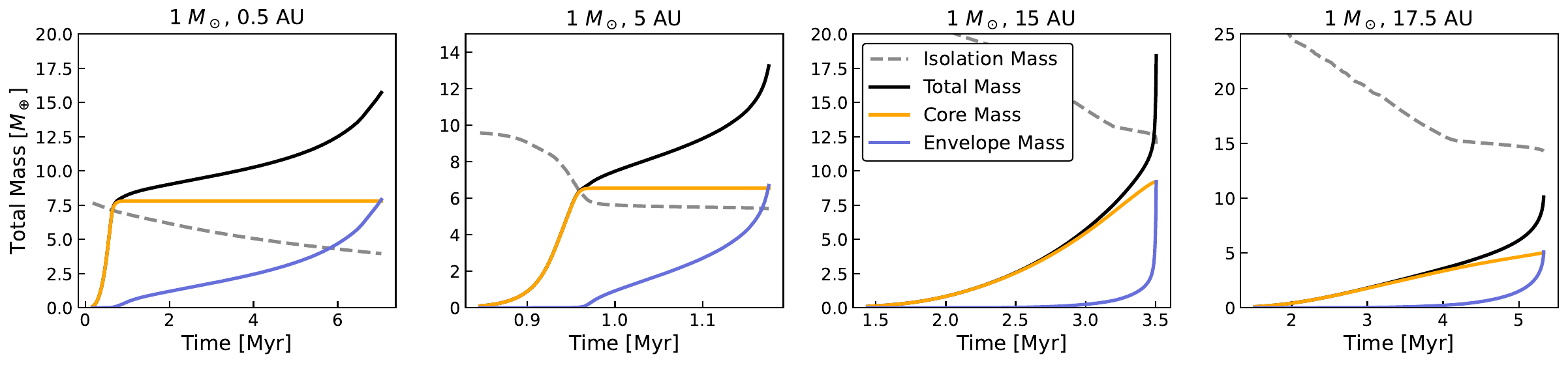} 
    \caption{Examples of four accretion tracks for planets forming around a 1 $\mathrm{M}_{\odot}$ star. The four panels show the  accretion until crossover ($M_{\rm core}$ = $M_{\rm envelope}$) for different starting (at 0.1 $\mathrm{M}_{\oplus}$) orbital distances, respectively at 0.5, 5, 15, and 17.5 AU. The evolving core mass is shown in orange, the envelope mass in blue, and the total mass ($M_{\rm core}$ + $M_{\rm envelope}$) is in black. The theoretical evolving line denoting isolation mass described by eq.~\ref{eq.isolation} is shown in dashed grey. The initial time of the simulation shown on the x axes is informed by the initial formation timescales at 0.1 $\mathrm{M}_{\rm \oplus}$ that can be seen in Figure~\ref{fig.initial_timescale}, and found in Section~\ref{sec:formation_timescales}. Note that the axis ranges are different for the different panels.} 
    \label{fig.4acc}
\end{figure*}
\subsection{Different regimes}
\label{Two_regimes}
Fig.~\ref{fig.4acc} shows four representative formation tracks around a  1 $\mathrm{M}_{\rm \odot}$ star,  illustrating the isolation mass transition. The plots show the accretion history of the core (heavy elements) and the envelope (hydrogen-helium) as a function of time. We also show  the total planetary mass (core+envelope). Also shown is the isolation mass (dashed grey line), which represents the theoretical mass at which the planet halts pebble accretion due to the formation of a partial cavity. This line decreases over time, reflecting the planet's inward migration via Type I migration and the disc's thermal cooling.

At 0.5 AU isolation mass is reached relatively quickly due to the small orbital separation. However, viscous heating increases  the isolation mass once the planets reach the radiation-dominated boundary. The core reaches $\sim$7 $\mathrm{M}_{\rm \oplus}$ at pebble isolation. This is massive enough to lead to giant planet formation. However, crossover is only reached after $\sim$7~Myr, roughly twice the adopted disc lifetime of 3.7~Myr for a solar-mass star (Eq.~\ref{eq:disc_lifetime}). Within our fiducial disc lifetime, gas accretion during phase~2 is therefore too slow for runaway gas accretion to occur, leaving the planet as a super-Earth or sub-Neptune. The thermodynamic origin of this prolonged cooling timescale is discussed in Section~\ref{Inner_disc}. 
\par

At 5~AU the classical pathway is recovered: rapid solid accretion drives the core's growth to reach isolation mass followed by crossover mass. The envelope contraction progresses more rapidly than at 0.5~AU ($\sim0.2$ versus $\sim6$ Myr after reaching pebble isolation mass), due to the lower initial disk temperature (${\sim}75$ versus ${\sim}637$~K). Since the final core masses of the 0.5 and 5 AU cases are comparable, it is clear that the ability to reach crossover is controlled by the disc's physical conditions and its lifetime.

At 15~AU the system enters a transitional regime: the envelope is massive enough that the total planetary mass (black line) reaches isolation mass (dashed-grey line) with the core mass being well below $M_{\rm iso}$. This would halt solid accretion at a sub-isolation core mass. Physically, the disc responds to the total gravitational mass when forming the pressure bump, independent of the planetary composition. Hence, pebble accretion is reduced when the total mass reaches the  isolation mass, but in this case phase~2 is  short because gas accretion already dominates the growth. Gas accretion continues efficiently, but the final crossover mass falls below $2M_\mathrm{iso}$ and formation timescales are extended. 

At 17.5~AU the picture is different: the planet reaches crossover mass before its core grows to the isolation mass. The pebble accretion rate remains low but roughly steady, so the core keeps accreting solids until  crossover. In some cases, such as with $0.5\, \mathrm{M}_{\rm \odot}$, the pebble accretion rate falls from a few$\,\times10^{-7}$ down to $\sim10^{-8}$--$10^{-9}\,\mathrm{M}_{\rm \oplus}\,{\rm yr}^{-1}$ at wide orbits. Because the core remains small, gas accretion is slow and takes longer, but crossover is still reached below the isolation mass. This also occurs in the case of 0.5\,$\mathrm{M}_{\rm \odot}$ at 6.75~AU as can be seen in  Fig.~\ref{fig.masses}.
\par 
The protoplanet that originally forms at 17.5 AU starts in a disc with a surrounding temperature of 44 K  (which increases due to inward migration), while a planet that forms at 0.5 AU is surrounded by a hot gas (initial temperature of 640 K) where viscous heating is significant. Although this exceeds the characteristic disc lifetime adopted in this study, it remains within the $1\sigma$ uncertainty of observationally inferred disc dispersal timescales. Consequently, this end-member scenario should not be regarded as unphysical, but rather as a low-probability yet physically plausible outcome lying in the tail of the disc lifetime distribution. 

\section{Discussion}
\label{Discussion}
\subsection{Inner disc: isolation and temperature controlled}
\label{Inner_disc} 

In the inner disc, giant planet formation is expected to be limited by the core mass, since the isolation mass is intrinsically low close to the star. However, viscous heating increases the disc’s temperature and inflates the scale height, especially at early stages (as this effects weakens with disc evolution). Since $M_{\rm iso}\propto(h/r)^3$ the isolation mass increases. Pebbles are still accreted in these locations although the flux interior to the ice line is reduced  (ice sublimation and the fragmentation of dry pebbles; see
Section~\ref{sec:Stokes} for discussion). Nevertheless, the core can still reach isolation mass.  In that case, the bottleneck is not reaching the core mass, but the cooling: the Kelvin--Helmholtz timescale after isolation exceeds the disc's lifetime due to the high disc temperatures controlled by viscous heating. This  prevents the envelope from contracting efficiently. 

Generally, the core mass affects gas accretion more strongly than the disc's  temperature,
as $\dot{M}_{\rm gas} \propto M_{\rm core}^{11/3}\,T^{-1/2}$ \citep{Bitsch_2015}.
Close to the star, however, this advantage disappears. The pebble isolation mass
scales as $M_{\rm iso} \propto (T r)^{3/2}$, so a planet closer to the star  can build a core comparable in mass to the case of a planet forming further out only  if the disc is much hotter: at 0.5~AU the disc would need 
to be ten times hotter than at 5~AU to reach the same isolation mass. 
The effect of viscous heating nearly creates such a condition, as the temperature further in increases enough for $M_{\rm iso}$ to flatten out in the inner disc (Fig.~\ref{fig.isolation_masses}).

To investigate which factor slows down the planet closer in, we split the ratio of the gas
accretion rates at the two orbits into a core-mass part and a temperature part right at the beginning of Phase~2:
\begin{equation}
\begin{split}
  \frac{\dot{M}_{\rm gas,\,0.5}}{\dot{M}_{\rm gas,\,5}}
  &= \left(\frac{M_{\rm core,\,0.5}}{M_{\rm core,\,5}}\right)^{11/3}
     \left(\frac{T_{5}}{T_{0.5}}\right)^{1/2} \\
  &= \left(\frac{7.1}{6.4}\right)^{11/3}
     \left(\frac{98\,{\rm K}}{936\,{\rm K}}\right)^{1/2}
   \simeq 1.5 \times 0.32 \simeq 0.5 .
\end{split}
\label{eq.split}
\end{equation}
where the subscripts refer to the planets formed at 0.5 and 5~AU, respectively, with the quantities evaluated when each planet reaches its isolation mass (by which point they have migrated to 0.33 and 2.9~AU, respectively). The more massive core favours faster formation for the planet originating at 0.5~AU. In contrast, the temperature term, with values of ${\sim}936$~K and ${\sim}98$~K, respectively, amounts to ${\simeq}\,0.32$ and acts in the opposite direction. Thus, it is not the smaller core that hinders formation at smaller orbital distances, but rather the higher disc temperature. The simulations reveal an even more pronounced effect than suggested by this simple estimate: the planet at 0.5~AU takes nearly 30 times longer to form, compared to a factor of only two predicted by the estimate.

Because the envelope is fully embedded in the disc, this high temperature imposes a thermal floor from which Kelvin--Helmholtz contraction must proceed, suppressing radiative cooling and lengthening $\tau_{\rm KH}$ beyond the disc lifetime. Inward migration amplifies the effect: as the planet drifts into hotter regions, the disc temperature at the envelope boundary can exceed the envelope's own outer temperature, reversing the heat flux and actively heating the envelope rather than allowing it to cool.
Viscous heating thus aids the assembly of more massive cores but, through the hotter disc it produces, ultimately impedes the cooling needed to reach crossover.

Neither Type~I nor Type~II migration appears capable of fully delivering planets into the hot-Jupiter region within our framework. Even under the optimistic migration track of \citet{Tanaka_2020} (Table~\ref{tab.typeII}), Jupiter-mass planets remain at orbital periods of tens of days. This estimate is only indicative, however, as it does not model growth and migration self-consistently; more detailed Type~II migration calculations are therefore required. Within these limitations, high-eccentricity migration \citep{Winn_2010, Dawson_2018, Fan_2026} remains the most plausible pathway for producing the shortest-period giant planets that are not reproduced by any model considered here.

\subsection{Outer disc: a second formation regime beyond isolation}
\label{Outer_disc}
At larger orbital distances, a qualitatively different regime emerges. In these cases, the planet reaches the pebble isolation threshold before the core does, as shown for the 15 and 17.5~AU cases around a 1~$\mathrm{M}_{\rm \odot}$ host in Fig.~\ref{fig.4acc}. In some extreme situations, inward pebble drift exhausts the local solid reservoir before isolation is  reached (recorded for the 0.25 $\mathrm{M}_{\rm \odot}$, 2.9 AU scenario). In either case, the core mass at crossover remains below the pebble isolation mass, in contrast to standard formation models that assume the core must reach the pebble isolation mass before runaway gas accretion can begin. This alternative pathway removes the need for a massive heavy-element core to initiate runaway accretion.  
\par 
\begin{figure*}[!b]
    \centering
    \includegraphics[width=1\textwidth]{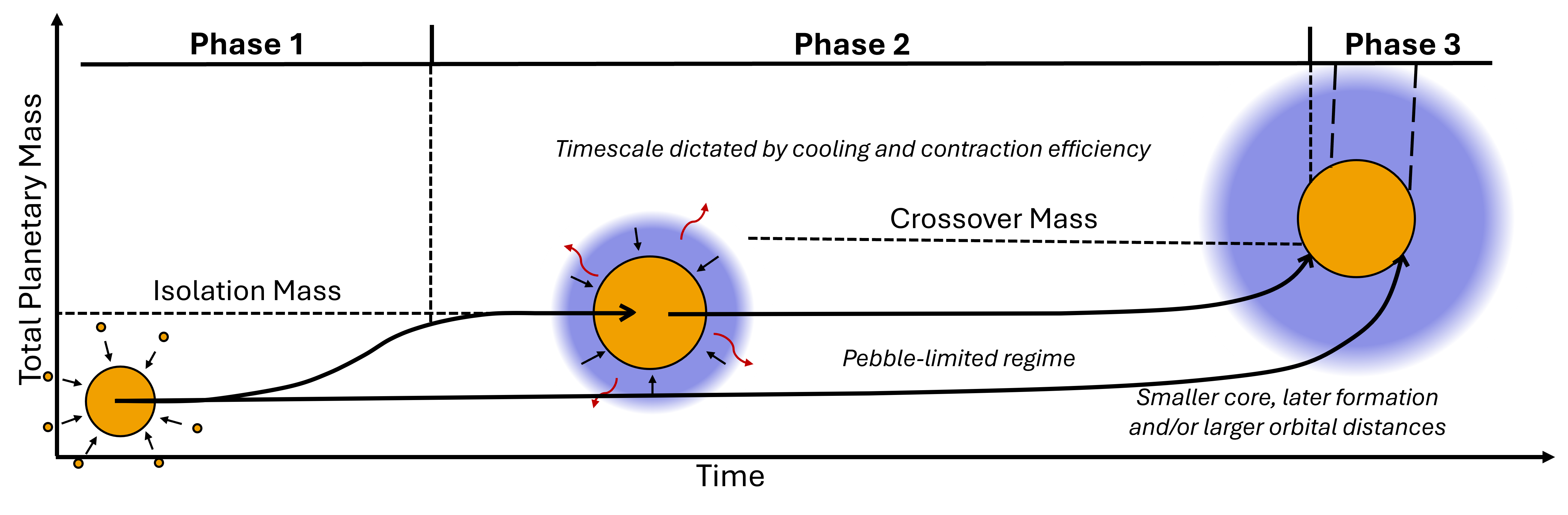} 
    \caption{A sketch showing the two pebble accretion pathways depending on the formation location. 
    The three standard phases that can be distinguished are a solid accretion dominated (phase 1), followed by the isolation mass, cooling-controlled phase (phase 2), to end with the gas-dominated phase 3. We find an  alternative path that does not require isolation mass physics to contract due to a lower core mass at higher orbital distances. The success in forming a planet stems in the cooling efficiency (controlled by the grain opacity $f_{\rm g}$) and the availability of pebbles and gas.} 
    \label{fig.diagram}
\end{figure*}

Runaway is reached for two reasons. First, once solid accretion ceases (whether by isolation or by exhaustion of the pebble supply), the accretion luminosity that supports the envelope against contraction vanishes, and the envelope can cool and contract. Even when pebble accretion continues, as in the 17.5~AU case, the accretion rate remains sufficiently low (on average $\dot{M}_{\rm Z} \sim
10^{-6}\,\mathrm{M}_\oplus\,\mathrm{yr}^{-1}$ over the late pre-crossover phase, for a core of $\sim 4\,\mathrm{M}_\oplus$) around 1~$\mathrm{M}_\odot$. If the pebble accretion is inefficient, the planet's ability to accrete is dictated by contraction. Envelope contraction is as a result efficient because of lower accretion luminosities given the smaller cores.
Furthermore, the cold, low-density outer disc reduces the crossover mass  \citep{Piso_2014}. Under these conditions, the envelope contracts at a much lower mass, enabling runaway gas accretion also for relatively small core masses. Therefore, we conclude that the onset of runaway gas accretion does not necessarily require a massive heavy-element core as often assumed. 
It is the core mass that determines the balance. A more massive core contracts its envelope more rapidly, corresponding to a shorter Kelvin–Helmholtz timescale \citep{Ikoma_2000, Bitsch_2015}. However, in this regime the core remains capped below the pebble isolation mass and cannot grow indefinitely. Conversely, if the core is too small, its Kelvin–Helmholtz timescale exceeds the disc lifetime, preventing the onset of runaway gas accretion. Therefore, successful giant planet formation in the outer disc requires a core that is sufficiently massive to contract its envelope within the disc lifetime, while residing in a disc that is cold and tenuous enough to facilitate runaway gas accretion. 

This second regime is not universal, as it requires specific disc conditions, orbital distances, and relatively low grain opacities that may not occur in all systems. This alternative channel has also been identified by \citet{Gurrutxaga_2024} who showed that smaller cores can lead to giant planet formation at larger separations due to the decay in pebble supply from the disc's reservoir. The qualitative agreement between our results and those of \citet{Gurrutxaga_2024} provides support for this alternative channel of giant planet formation.

The existence of this second regime does not make giant planet formation more efficient at large separations but rather extends the outer formation location for giant planets. Runaway gas accretion can occur for planetary cores below the pebble isolation mass. The predominant outcome at large orbital distances is planets that fail to undergo envelope contraction before the gas disc disperses, resulting in ice giant or sub-Neptunes. The existence of another regime suggest that pebble accretion theory accommodates a broader diversity of formation pathways than the standard picture assumes. A schematic of the two pathways is presented in Figure~\ref{fig.diagram}. 

\subsection{Diversity in core masses}
\label{Core_diversity}
The inferred core masses in this study are between ${\sim}1$ and
$21\,\mathrm{M}_{\rm \oplus}$ ($1.05$--$20.59\,\mathrm{M}_{\rm \oplus}$ for in-situ and $1.79$ -$17.74\,\mathrm{M}_{\rm \oplus}$ for migrating planets). 
The narrower range for migrating planets reflects two effects. The smallest cores form at large radial distances, where the pebble isolation mass is never reached. As they migrate inward, they accrete more efficiently and can grow larger, raising the lower bound from $1.05$ to
$1.79\,\mathrm{M}_\oplus$. At smaller radial distances, cores with similar masses do not lead to the formation of giant planets. This demonstrates the effect of the disc's temperature on the  planetary cooling and contraction. Migration lowers $M_{\rm iso}$, reducing the upper bound from $20.59$ to $17.74\,\mathrm{M}_\oplus$. 

This range overlaps with classical estimates of the critical core mass $M_{\rm crit}$ from \citet{Stevenson_1982}, \citet{Mizuno_1980}, and \citet{Pollack_1996}, although these estimates were derived in a planetesimal accretion context and refer to the minimum core mass required to trigger runaway gas accretion. In both frameworks, it is ultimately the core mass that determines whether a planet becomes a gas giant, making the comparison meaningful. Indeed, \citet{Ikoma_2000} showed that $M_{\rm crit}$ can span ${\sim}0.8$--$20\,\mathrm{M}_{\rm \oplus}$ depending on disc opacity and accretion conditions, consistent with the range inferred here when using the stellar mass and orbital distance as variables. Consistently, \citet{Piso_2014} found that the critical core mass decreases with orbital distance ($M_\mathrm{crit} \propto a^{-0.3}$), primarily due to the lower disc temperatures in the outer disc. This trend of decreasing critical mass with increasing orbital separation is consistent with the low-mass cores reaching crossover at large orbital distances in our simulations. 

Beyond this numerical agreement, the simulations reveal an additional structure. Small cores arise not only in the inner disc, where $M_{\rm iso}$ is intrinsically low, but also at large orbital distances, where inward pebble drift depletes the solid reservoir before gap opening occurs. 
In this outer pebble-limited regime, the core never reaches $M_{\rm iso}$, yet runaway accretion can still occur if the envelope cools efficiently and disc gas remains available long enough. The key physical balance is between the heating from the infalling solids, the radiative cooling of the envelope set by its opacity, and the availability of nebular gas before disc dispersal. This is consistent with the 
$\dot{M}_{\rm core}\,\kappa$ dependence identified by \citet{Stevenson_1982}, and suggests that runaway gas accretion does not require a unique core mass threshold but can occur under a broad range of conditions. We therefore argue that the traditional concept of a critical core mass of $\sim$10\,M$_{\oplus}$ for giant planet formation should be abandoned. 

The resulting core mass diversity has direct implications for the compositions of giant and intermediate-mass planets. Planets forming where pebble isolation is reached tend to have larger cores and higher heavy-element fractions, while those forming in pebble-limited environments can undergo runaway accretion with much smaller cores. Two giant planets with the same final mass can therefore have very different bulk compositions (and internal structures), depending on where and how they formed.  This connects naturally to the findings of \citet{Muller_2025}, who show that giant planets around low-mass stars tend to be depleted in heavy elements, a trend our model reproduces: lower stellar masses reduce $M_{\rm iso}$, while longer disc lifetimes around such stars \citep{Ribas_2015} give more time for envelope cooling and gas accretion to proceed and deliver gas giants with lower-mass cores. We note, however, that \citet{Chachan_2025} did not find any heavy element depletion for planets around low-mass stars and that this topic is still being investigated. In any case, this effect is likely to be obscured by the stochastic variations in the parameters held fixed in this study. The diversity of cores retrieved for each stellar host already suggests substantial variability that may not be statistically interpretable in population-level analyses. 
More broadly, the observed spread in giant planet metallicities may encode information about formation environment. Stellar mass, orbital location, solid supply, and disc lifetime all influence the core mass at which runaway accretion begins, producing a range of heavy-element enrichments rather than a single mass-metallicity relation. This may also explain the low-metallicity tail of the observed giant planet population \citep{Thorngren_2016, Muller_2025, Chachan_2025}. The trends identified here offer a potential link between giant planet compositions and the physical conditions of their formation. 

\subsection{Diversity in internal structure and bulk composition}
\label{Bulk_composition}

The formation history determines both the planet's bulk composition and its internal structure. 
Our results clearly show that the planetary bulk composition depends on the stellar mass, orbital distance, and migration history. If we consider the scenario in which planets stop accreting at indicative mass cutoffs, we can compare the inferred bulk composition for the different cases. Table~\ref{tab:bulk_composition} lists the inferred bulk metallicity $Z = M_{\rm core}/M_{\rm tot}$ at four mass thresholds (2.5, 5, 10, and 20 $\mathrm{M}_{\rm \oplus}$) for each combination of stellar mass and orbital distance for our migrating simulations.  
We note that final planetary composition can change due to enrichment during runaway and post-formation would also affect the final bulk composition. 
Indeed, processes such as planetesimal accretion \citep{Alibert_2018, Shibata_2019, Shibata_2020, Shibata_2022,  Turrini_2021, Danti_2023, Shibata_2024}, accretion of enriched disk gas \citep{Booth_2017, Booth_2019, Schneider_2021a, Schneider_2021b}, giant impacts \citep{Ogihara_2021, Gabriel_2023}, and atmospheric retention of ablated solids \citep{Brouwers_2020} can change the planetary  composition and internal structure. 

\begin{figure}[!htb]
    \centering
    \includegraphics[width=1\columnwidth]{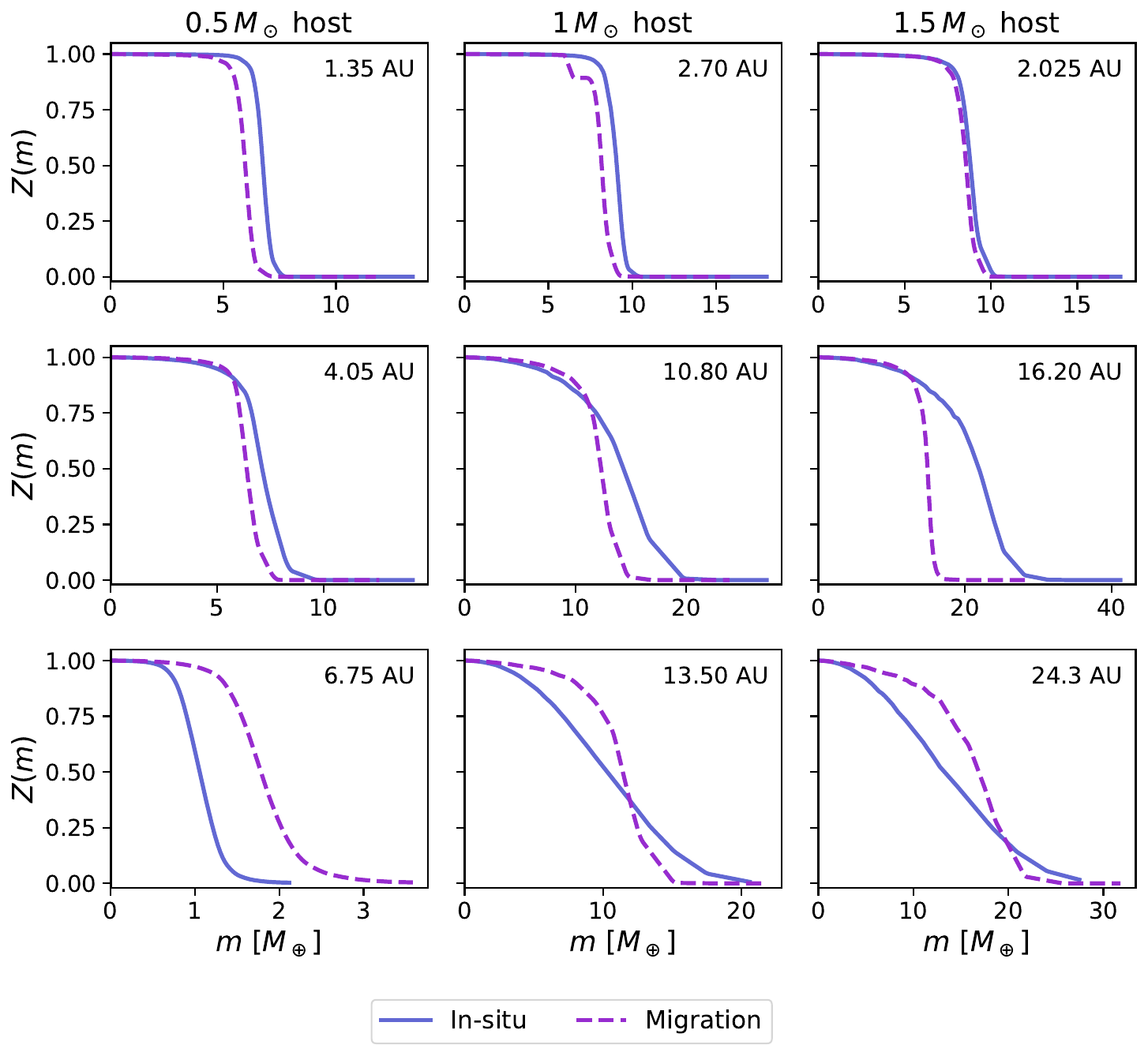} 
    \caption{Inferred heavy-element profiles as a function of cumulative planetary mass $m$ for selected in-situ (blue) and migrating (dashed purple) simulations. Each panel corresponds to a representative orbital distance (labelled) for host stellar masses of 0.5, 1, and $1.5\,\mathrm{M}_{\rm \odot}$ (see text for details). } 
    \label{fig.3x3}
\end{figure}

In addition, it has been shown that the ratio between the solid and gas accretions determines the composition gradient within the planetary interior and its steepness \citep{Helled_Stevenson_2017,Valletta_2020}.  
The heavy-element distribution within the planet $Z(m)$, assuming that convective mixing is negligible during the planetary growth, can be approximated by \citep{Helled2022}: 
\begin{equation}
Z(m) \simeq \left.\frac{\dot{M}_{\rm Z}}{\dot{M}_{\rm tot}}\right|_{M(t)\,=\,m}
\label{eq.comp}
\end{equation}
where $\dot{M}_{\rm Z}$ and $\dot{M}_{\rm tot}$ are the solid (heavy elements) and total (gas + solids) accretion rates at each mass coordinate $m$, and $t$ is time. 

We take $\dot{M}_{\rm Z}$ and $\dot{M}_{\rm tot}$ from our simulations and reconstruct $Z(m)$ up to the crossover mass. We select representative runs around the $0.5$, $1$, and $1.5\,\mathrm{M}_{\rm \odot}$
hosts, sampling three orbital distances per star: a close-in case, the case reaching the highest crossover mass, and an outer case (see the accretion tracks in Fig.~\ref{fig.masses}). Fig.~\ref{fig.3x3} shows the heavy-element distributions obtained from the in-situ and
migrating simulations. The inferred heavy-element distributions are determined by the accretion rates and are shown for a given final planetary mass from the planetary centre outward (shown in Eq.~\ref{eq.comp}), where $Z = 1$ corresponds to pure heavy elements and $Z = 0$ to a pure hydrogen-helium mixture.  

The inferred profiles reveal distinct accretion regimes that lead to different predicted internal structures. At smaller orbital distances (top row, Fig.~\ref{fig.3x3}), despite differences in the final crossover mass, planets have  similar structures across different stellar hosts with a structure that resembles a core-envelope structure followed by a small heavy-element gradient that decays rather quickly.  
At larger orbital distances, the core does not reach isolation mass at crossover. Pebble accretion continues (since isolation mass is not a limiting factor) leading to heavy-element enrichment further from the planetary centre. The resulting structure has heavy-element gradients that are more extended.  In the most extreme cases (bottom row), inward migration allows the protoplanets to accrete pebbles over a longer trajectory through the disc, further extending the compositional gradients and potentially explaining the formation of ``fuzzy cores''. In these cases the planets also tend to have higher metallicities as listed in Table~\ref{tab:bulk_composition}. 

\begin{table}[htbp]
\centering
{\small
\caption{Bulk metallicity at fixed planet mass cutoffs for simulations with Type-I migration.}
\label{tab:bulk_composition}
\begin{tabular}{|c|c||c|c|c|c|}
\hline
\multicolumn{2}{c||}{\textbf{Formation conditions}} & \multicolumn{4}{c}{\textbf{Planetary metallicity} $Z$ \textbf{at mass cutoff}} \\
\hline
$\boldsymbol{M_\star}$ [$\boldsymbol{\mathrm{M}_{\rm \odot}}$] & $\boldsymbol{a_{0}}$ \textbf{[AU]} & $\boldsymbol{2.5\,\mathrm{M}_{\rm \oplus}}$ & $\boldsymbol{5\,\mathrm{M}_{\rm \oplus}}$ & $\boldsymbol{10\,\mathrm{M}_{\rm \oplus}}$ & $\boldsymbol{20\,\mathrm{M}_{\rm \oplus}}$ \\
\hline\hline
0.25 & 1.45 & 0.999 & 0.836 & -- & -- \\
0.25 & 2.175 & 0.928 & -- & -- & -- \\
\hline
0.5 & 0.675 & 0.999 & 0.984 & 0.503 & -- \\
0.5 & 1.35 & 0.999 & 0.995 & 0.591 & -- \\
0.5 & 2.7 & 0.999 & 0.997 & 0.630 & -- \\
0.5 & 4.05 & 0.998 & 0.992 & 0.632 & -- \\
0.5 & 5.4 & 0.992 & 0.949 & 0.533 & -- \\
0.5 & 6.75 & 0.712 & -- & -- & -- \\
\hline
1 & 1.35 & 1.000 & 0.998 & 0.757 & -- \\
1 & 2.7 & 1.000 & 0.999 & 0.808 & -- \\
1 & 5.4 & 1.000 & 0.999 & 0.870 & -- \\
1 & 8.1 & 0.999 & 0.997 & 0.969 & -- \\
1 & 10.8 & 0.998 & 0.994 & 0.972 & 0.596 \\
1 & 13.5 & 0.996 & 0.987 & 0.942 & 0.546 \\
\hline
1.5 & 1 & 1.000 & 0.999 & 0.898 & -- \\
1.5 & 2.025 & 0.999 & 0.998 & 0.842 & -- \\
1.5 & 4.05 & 1.000 & 0.999 & 0.934 & -- \\
1.5 & 8.1 & 1.000 & 0.999 & 0.992 & 0.546 \\
1.5 & 12.15 & 0.999 & 0.998 & 0.992 & 0.611 \\
1.5 & 16.2 & 0.999 & 0.997 & 0.989 & 0.721 \\
1.5 & 20.25 & 0.998 & 0.994 & 0.978 & 0.865 \\
1.5 & 24.3 & 0.996 & 0.987 & 0.962 & 0.787 \\
\hline
\end{tabular}
\tablefoot{$Z = M_{\rm core}/M_{\rm tot}$ is evaluated along each formation
track at the quoted mass, i.e.\ the inferred planetary composition if
accretion were truncated at that point. These values correspond to a stage where the 
envelope is still embedded in the disc, before post-formation contraction and
atmospheric mass loss, and are therefore not direct predictions of final observed
compositions. However, they suggest a possible trend. Note that $Z = 0.5$ corresponds to crossover by definition, so values approaching 0.5 correspond to cases where the planets are close to runaway accretion (at that mass).  Dashes indicate that the planet reached crossover mass below the quoted cutoff.}}
\end{table}

We note that in this study the accreted heavy elements were assumed to reach the planetary centre. Therefore, the deposition of heavy elements in the gaseous envelope and its effect on the planetary growth was not included. In reality, the deposition of heavy elements in the atmosphere can alter planetary growth, thereby affecting the subsequent accretion rate. However, the details of this process are complex and also depend on the assumed chemical composition of the accreted heavy elements and their sizes \cite{Valletta_2019,Valletta_2020,MolLous_2024}. Although the inferred heavy-element distribution we obtained may be oversimplified, it clearly shows the diversity of the internal structures and how they vary depending on the formation conditions. The inferred diversity in bulk compositions and internal structure naturally explains the large ranges of radii and masses of intermediate-mass planets. 
Finally, we note that the heavy-element distribution is also a key factor in determining the long-term thermal evolution of giant intermediate-mass planets. In particular, steep compositional gradients can suppress convection and modify heat transport within the planetary interior. This, in turn, influences the planet's contraction and, consequently, its observable properties and characterisation \citep{Knierim2025, Eberlein2025}.

\subsection{Limitations and Future Work}
Although our work represents a step forward in our understanding of planet formation, it also has limitations and motivates several avenues for future research. A key outstanding question is the physical origin of the observed pile-up of hot Jupiters at orbital periods of approximately 3 days and whether this feature reflects an intrinsic outcome of planet formation or is primarily the result of observational biases. It also remains to be established under which conditions Hill-sphere unbinding effectively suppresses gas giant formation and how strongly the pebble flux must decline for Hill-sphere shrinkage to dominate over gas accretion.

Another important open question concerns the role of the planetary core during gas giant formation. It is currently unclear whether a physically motivated minimum or maximum core mass exists, or whether the diversity of formation pathways makes such a threshold fundamentally degenerate. In this context, it would be important to determine whether gas giants formed in the outer disc with relatively small cores can be observationally distinguished from their inner-disc counterparts with larger cores, for example through measurements of their bulk heavy-element content or atmospheric metallicities with facilities such as JWST and Ariel.

The dependence of planet formation on disc microphysics also deserves further investigation. In particular, the sensitivity of the outer-disc formation regime to the assumed grain opacity  and the physical processes that determine its value in realistic protoplanetary discs remain poorly constrained.

Also, as discussed above, the deposition of heavy elements within the planetary envelope and atmosphere should be treated self-consistently, as it can modify the envelope's opacity, mean molecular weight,  thermal structure and therefore the planetary growth. These effects influence the cooling and contraction of the planet, regulate the gas accretion rate, and ultimately shape the planet's final composition and internal structure. Incorporating these processes is crucial for predicting observable quantities, such as atmospheric metallicities and bulk heavy-element fractions. Also, extending the present models beyond the onset of runaway gas accretion is essential for connecting formation models with the observed exoplanet population. Future studies should follow the growth of planets through runaway accretion, including the accretion and redistribution of heavy elements after crossover, to determine their final masses, compositions, and internal structures.

Finally, our models rely on scaling relations for protoplanetary disc properties that remain uncertain and may not accurately represent the true diversity of disc properties. Improved observational constraints on the distributions of disc lifetimes, masses, and other key physical properties across different stellar host masses is therefore essential for reducing systematic uncertainties in planet formation models and testing the robustness of their predictions. Improved measurements of protoplanetary discs together with detections and characterisation of planetary systems around different stellar types, will provide valuable tests of competing planet formation scenarios.

\section{Summary and Conclusions}
\label{sec:conclusion}

In this study, we simulated planet formation by pebble accretion around different stellar types accounting for various formation conditions. We showed that the resulting planetary mass, composition, and structure strongly depend on the formation conditions. Our main findings can be summarised as follows:

\begin{itemize}

    \item \textit{Embryo formation timescales are the primary bottleneck.} The initial growth of a planetary embryo to a size capable of efficient pebble accretion sets the pace for all subsequent evolution, and represents the most critical constraint on successful gas giant formation.
\medskip
    \item \textit{In-situ hot Jupiter formation is strongly suppressed.} The elevated temperatures of the inner disc lead to extended Kelvin--Helmholtz contraction timescales $\tau_{\rm KH}$, making in-situ formation of hot Jupiters difficult to reconcile within this framework. Neither Type~I nor the Type~II prescription adopted here reproduces the short orbital periods of the observed hot Jupiter population, although other migration treatments may do so.
\medskip
    \item \textit{Migration inhibits gas giant formation around low-mass stars.} Around M-dwarf hosts ($0.1$--$0.5\,\mathrm{M}_{\rm \odot}$), inward migration can keep protoplanets interior to the water ice line, where their cores stall below the pebble isolation mass, and at later times drives Hill sphere contraction faster than the envelope can be replenished by gas accretion, due to lower local gas surface densities. Combined with longer embryo formation timescales and more hostile disc environments, this significantly complicates gas giant formation around low-mass stars, and may contribute to the observed paucity of hot Jupiters in this stellar mass regime, though a more detailed investigation is required.
\medskip
    \item \textit{Close-in giant planet formation is limited by the ability of the envelope to cool and not by the core mass.}
    In the inner disc, pebble accretion efficiently brings the core to the local pebble isolation mass, which viscous heating raises to values comparable to those further out. However, the same high disc temperatures compress the envelope and suppress its cooling, so the planet cannot reach runaway gas accretion within the disc's lifetime.
\medskip
    \item \textit{A distinct formation regime exists in the outer disc.} Beyond a characteristic orbital radius, late embryo formation precludes reaching the isolation mass via pebble accretion. In this regime, whether a planet ultimately undergoes runaway gas accretion is governed by the efficiency of radiative cooling and the disc dissipation timescale.
\medskip
    \item \textit{A broad range of core masses is produced.} Under the disc conditions adopted here, the inferred masses of gas giants cores range between $1$ and $21\,\mathrm{M}_{\rm \oplus}$,  with the upper end set by the isolation mass in the inner disc and the lower end by the pebble supply in the outer disc. This shows that  giant planets can also form with very small cores and can have a large range of metallicities. 
    \medskip
\item \textit{The formation conditions (stellar host, formation location, migration) directly affect the planetary bulk composition and internal structure.} For a given planetary mass, the bulk metallicity and the predicted distribution of heavy elements can vary substantially, leading to the wide diversity of inferred exoplanet compositions. 
\end{itemize}

Overall, our results demonstrate that giant planet formation through pebble accretion is not governed by a single characteristic pathway, but by the complex interplay between embryo formation, orbital location, disc's properties, migration, and the mass of the host star. These factors determine not only whether a planet reaches runaway gas accretion, but also the composition  and internal structure of the resulting planet, naturally producing the remarkable diversity observed among giant exoplanets. The broad range of crossover core masses predicted by our simulations further suggests that giant planets retain a lasting imprint of their formation environment, with important consequences for their internal structure, bulk composition, and thermal evolution.  

Looking ahead, the rapidly advancing observational landscape offers an unprecedented opportunity to test these theoretical predictions. High-resolution observations of protoplanetary discs, increasingly precise measurements of exoplanet demographics across stellar populations, and detailed characterisation of giant planet interiors using upcoming data will place tighter constraints on the physical processes governing planet formation. By linking formation pathways to observable planetary properties, models such as those presented here can become increasingly predictive, allowing competing formation scenarios to be distinguished with growing confidence. We are entering an era in which theory and observations are converging to reveal not only how giant planets form, but why planetary systems exhibit such extraordinary diversity, a prospect that promises to transform our understanding of planetary origins. 

\begin{acknowledgements}
We thank the anonymous referee for a thorough and constructive report. We thank Henrik Knierim, Gabriele Cugno, Simon Müller, Dori Blakely, Saskia Hekker, Bertram Bitsch, Nerea Gurrutxaga, and David Paul Bennett for insightful discussions. We believe in open and accessible science, and this work is freely available as an open-access publication.
\end{acknowledgements}

\bibliographystyle{aasjournal}
\bibliography{biblio}

\begin{appendix}
\nolinenumbers
\section{Disc model formulation}
\label{appendix:disc_model}

\subsection{Thermal structure}

The disc midplane temperature is determined by the combination of stellar irradiation and viscous heating.

The irradiation temperature is given by

\begin{align}
T_{\mathrm{irr}} = 150\,\mathrm{K}
\left(\frac{L_{\rm \star}}{\mathrm{L}_{\rm \odot}}\right)^{2/7}
\left(\frac{M_{\rm \star}}{\mathrm{M}_{\rm \odot}}\right)^{-1/7}
\left(\frac{r}{1\,\mathrm{AU}}\right)^{-3/7},
\label{eq:T_irr}
\end{align}
following \citet{Shibata_2025}.

The viscous temperature follows from the vertically integrated dissipation rate \citep{Pringle_1981}:

\begin{align}
T_{\mathrm{visc}} = \left(\frac{3GM_\star\dot{M}}{8\pi\sigma_{\mathrm{SB}} r^3}\right)^{1/4},
\label{eq:T_visc}
\end{align}
where $\dot{M}$ is the local disc accretion rate (Eq.~\ref{eq:Mdot}).
The midplane temperature is then obtained by accounting for vertical radiative transfer through the disc column \citep{Nakamoto_1994},

\begin{align}
T_{\mathrm{mid}}^4 = T_{\mathrm{irr}}^4 + \left(\frac{3\tau}{8} + \frac{1}{2\tau}\right)T_{\mathrm{visc}}^4,
\label{eq:T_mid}
\end{align}
where $\tau = \kappa(T_\mathrm{mid})\,\Sigma_{\rm gas}/2$ is the optical depth to the
midplane (half the total column), with the Rosseland mean opacity
$\kappa = 0.1\,(T/{\rm K})^{0.5}$~cm$^2$\,g$^{-1}$ \citep{Bell_1994}.
Above ${\sim}1500$~K dust sublimates, so we remove the dust opacity there; this
only affects the innermost disc ($r\lesssim0.1$~au). When the disc is optically thick
($\tau \gg 1$), viscous heat is trapped and $T_{\mathrm{mid}} \gg T_{\mathrm{visc}}$.
Since $\kappa$ and $\Sigma_{\rm gas}$ depend on $T_{\mathrm{mid}}$,
Eq.~(\ref{eq:T_mid}) is solved iteratively together with Eq.~(\ref{eq:Sigma_gas}),
taking its lowest root.

\subsection{Disc accretion rate and gas surface density}

The disc evolves following the self-similar viscous solution of
\citet{Lynden-Pringle_1974}, computed with the viscosity of the irradiated disc,
$\nu_{\rm irr} = \alpha_{\rm acc} H_{\rm gas}^2(T_{\rm irr})\,\Omega_{\rm K}$, where
$\alpha_{\rm acc} = 10^{-3}$ is the viscosity parameter, $H_{\rm gas} = c_{\rm s}/\Omega_{\rm K}$
the disc scale height, $c_{\rm s}$ the sound speed and $\Omega_{\rm K}$ the Keplerian
frequency. Since $T_{\rm irr} \propto r^{-3/7}$, this viscosity scales as
$\nu_{\rm irr} \propto r^{\gamma}$ with $\gamma = 3/2 + {\rm d}\ln T_{\rm irr}/{\rm d}\ln r = 15/14$.
The local disc accretion rate is taken as the steady-state value
$\dot{M} = 3\pi\nu_{\rm irr}\Sigma$ of this solution, which gives
\begin{align}
\dot{M}(r,t) = \frac{M_{\mathrm{disc},0}}{2(2-\gamma)\,\tau_{\mathrm{vis}}}\,
  \mathcal{T}^{-\frac{5/2-\gamma}{2-\gamma}}
  \exp\left[-\frac{1}{\mathcal{T}}\left(\frac{r}{R_{\mathrm{disc}}}\right)^{2-\gamma}\right]
  \frac{1}{1 + \left(r_{\mathrm{mag}}/r\right)^4},
\label{eq:Mdot}
\end{align}

where $M_{\mathrm{disc},0}$ and $R_{\mathrm{disc}}$ are the initial disc mass and
the disc radius (Appendix~\ref{sec:scaling}), $\mathcal{T} = 1 + t/\tau_{\mathrm{vis}}$,
with $t$ the disc age and $\tau_{\mathrm{vis}}$ the viscous timescale given below, and
$r_{\rm mag} = 0.05$~au is the radius of the magnetospheric cavity, inside which the
disc is truncated. The viscous timescale is
\begin{align}
\tau_{\mathrm{vis}} = \frac{R_{\mathrm{disc}}^2}{3(2-\gamma)^2\,\nu_{\rm irr}(R_{\mathrm{disc}})}.
\label{eq:tau_vis}
\end{align}
The last factor in Eq.~(\ref{eq:Mdot}) suppresses the disc inside the magnetospheric truncation radius \citep{Koenigl_1991}, fixed at $r_{\mathrm{mag}} = 0.05$~au, the typical magnetospheric cavity radius of T~Tauri stars \citep{Bouvier_2007}.

The gas surface density follows from the steady-state relation \citep{Pringle_1981},
\begin{align}
\Sigma_{\mathrm{gas}} = \frac{\dot{M}}{3\pi\,\nu(T_{\mathrm{mid}})}, \qquad
\nu(T_{\mathrm{mid}}) = \alpha_{\mathrm{acc}}\,H_{\mathrm{gas}}^2(T_{\mathrm{mid}})\,\Omega_{\rm K}.
\label{eq:Sigma_gas}
\end{align}
The accretion rate is thus the same in the viscously heated and in the
irradiation-dominated parts of the disc: a hotter, more viscous region carries it with a
lower surface density. Far from the star, $T_{\mathrm{mid}} \to T_{\mathrm{irr}}$ and
$\Sigma_{\rm gas}$ reduces to the self-similar profile. Viscous heating is therefore
treated locally, at fixed $\dot{M}$, and does not feed back on the global disc
evolution.

\subsection{Scaling relations}
\label{sec:scaling}

We adopt scaling relations linking stellar and disc properties:

\begin{align}
L_{\star} &= 1\,\mathrm{L}_\odot
  \left(\frac{M_{\star}}{\mathrm{M}_\odot}\right)^{2}, \\
M_{\mathrm{disc},0} &= 0.1\,\mathrm{M}_\odot
  \left(\frac{M_{\star}}{\mathrm{M}_\odot}\right), \\
R_{\mathrm{disc}} &= 100\,\mathrm{au}
  \left(\frac{M_{\mathrm{disc},0}}{0.1\,\mathrm{M}_\odot}\right)^{0.625},
  \label{eq.rdisc}
\end{align}
following scaling laws provided by the literature \citep{Miguel_2019, Burn_2021, Chachan_2023, Venturini_2024}, and the fitted coefficients informed by observations
\citep{Choi_2016, Ramirez_2014, Andrews_2010, Andrews_2013}. These give $R_{\mathrm{disc}} = 23.7$--$128.8$~au over the stellar mass range we considered. Since the disc accretion rate and surface density declines exponentially beyond $R_{\mathrm{disc}}$, this radius also sets the duration of the inward pebble flux (Appendix~\ref{sec:accretion}). We note that the observational support for
Eq.~\ref{eq.rdisc} is weaker than for the mass scaling: homogeneous analyses of continuum and CO disc sizes find no statistically robust correlation between disc size and stellar mass \citep{Hendler_2020, Long_2022}. A weaker dependence would alter the duration of the pebble supply around low-mass stars.

This framework captures how the disc structure, temperature and surface density depend on stellar mass and evolve with time.

\section{Pebble Accretion Physics}
\label{sec:accretion}

The radial velocity of pebbles is given by \citep[e.g.,][]{Lambrechts_2014}:
\begin{align}
v_{\rm peb} = \frac{2 \eta v_{\rm K}}{\tau_f + \tau_f^{-1}} + \frac{\nu}{r},
\end{align}
where $\tau_f$ is the Stokes number, $v_{\rm K}$ is the Keplerian velocity, $\nu$ is the gas
viscosity, and $\eta$ quantifies the deviation from Keplerian rotation due to radial pressure
support. Evaluated at the disc midplane from the local aspect ratio and pressure gradient,
\begin{align}
\eta = \frac{1}{2} \left( \frac{H_{\rm gas}}{r} \right)^2
\left( \frac{3}{2} + \gamma - \frac{\partial \ln c_s}{\partial \ln r} \right),
\end{align}
equivalent to $\eta = -\tfrac{1}{2}(H_{\rm gas}/r)^2\,\partial\ln P_{\rm disc}/\partial\ln r$
with the pressure gradient decomposed into the gas surface-density and temperature
(sound-speed) slopes, where for the surface-density slope we use the self-similar
value $-\partial\ln\Sigma_{\rm gas}/\partial\ln r = \gamma$ (Appendix~\ref{appendix:disc_model}).
As such, $\eta$ varies with orbital distance and stellar mass through the disc structure.

The inward pebble flux is then expressed as
\begin{align}
\dot{M}_{\rm peb} = 2 \pi r \, \Sigma_{\rm peb} \, v_{\rm peb},
\end{align}
where $\Sigma_{\rm peb}$ is the pebble surface density. Following \citet{Lambrechts_2014}, the pebble surface density can be written as
\begin{align}
\Sigma_{\rm peb} = \frac{\dot{M}_{\rm peb}}{2 \pi r v_{\rm peb}}.
\end{align}

The pebble accretion rate onto a planetary core is given by \citep{Johansen_2017, Lambrechts_2014}:
\begin{align}
\dot{M}_{\rm core} = \pi R_{\rm peb,acc}^2 \, \rho_{\rm p,mid} \, \bar{S} \, \delta v,
\end{align}
where $R_{\rm peb,acc}$ is the pebble accretion radius, $\rho_{\rm p,mid}$ is the midplane pebble density, $\bar{S}$ is the stratification integral of pebbles, and $\delta v$ is the relative velocity between the planet and pebbles.

The midplane pebble density is given by
\begin{align}
\rho_{\rm p,mid} = \frac{\Sigma_{\rm peb}}{\sqrt{2\pi} H_{\rm peb}},
\end{align}
where the pebble scale height is
\begin{align}
H_{\rm peb} = H_{\rm gas} \sqrt{\frac{\alpha_{\rm turb}}{\tau_f}}.
\end{align}

The relative velocity between pebbles and the planet is approximated as
\begin{align}
\delta v = \Delta v + \Omega_K R_{\rm peb,acc},
\end{align}
where $\Delta v = \eta v_{\rm K}$ is the sub-Keplerian velocity.

The accretion radius $R_{\rm peb,acc}$ is determined by solving
\begin{align}
\tau_f = \Omega_{\rm K}\,\frac{\xi_B \Delta v + \xi_H \Omega_K R_{\rm peb,acc}}{G M_p / R_{\rm peb,acc}^2},
\end{align}
where $M_p$ is the planetary mass and $\xi_B$ and $\xi_H$ $= 0.25$ are fitting parameters that smoothly connect the Bondi and Hill accretion regimes \citep{Ormel_2010, Lambrechts_2012}.

\subsection{Stokes number and the water ice line}
\label{sec:Stokes}
The Stokes number $\tau_f$ and the pebble surface density depend on whether the planet lies interior or exterior to the water ice line, which we place at $T_{\rm disc}=170$~K \citep{Lambrechts_2014}.
Exterior to the ice line, pebbles retain their icy mantles and their size is set by the drift--coagulation balance. Following \citet{Lambrechts_2014}, the pebble surface density is given by:
\begin{align}
\Sigma_{\rm peb} = \sqrt{\frac{2\,\dot{M}_{\rm peb}\,\Sigma_{\rm gas}}
{\sqrt{3}\,\pi\,\epsilon_P\,r\,v_{\rm K}}},
\end{align}
and the associated Stokes number is:
\begin{align}
\tau_f = \frac{\sqrt{3}\,\epsilon_P\,\Sigma_{\rm peb}}{8\,\eta\,\Sigma_{\rm gas}},
\end{align}
with a coagulation efficiency $\epsilon_P = 0.5$.

Interior to the ice line ($T_{\rm disc}>170$~K), the water ice sublimates: we then reduce the pebble mass flux by a factor of two (assuming a water mass fraction of $0.5$) and treat the
surviving refractory grains as compact silicates of a fixed radius $R_{\rm peb}=0.1$~cm and internal density $\rho_\bullet = 5.5\,\mathrm{g\,cm^{-3}}$. Their Stokes number then follows from the physical grain size:
\begin{align}
\tau_f = \frac{\rho_\bullet\,R_{\rm peb}}{\rho_{\rm gas}\,H_{\rm gas}},
\end{align}
where $\rho_{\rm gas}$ and $H_{\rm gas}$ are the midplane gas density and the scale height, respectively. Both
the halved flux and the smaller, millimetre-sized silicate pebbles act to suppress the solid
accretion rate interior to the ice line.

\section{Type I migration}
\label{appendix:migration}

Low-mass planets embedded in a gaseous disc undergo Type I migration due to gravitational torques exerted by the surrounding gas \citep{Tanaka_2002, Paardekooper_2011}. These torques arise from two main contributions: Lindblad torques, generated by spiral density waves launched at Lindblad resonances, and corotation torques, associated with gas in the co-orbital region.

The radial migration rate can be written as

\begin{align}
\frac{{\rm d}r}{{\rm d}t} = \frac{2 \Gamma}{M_{\rm p} \Omega_{\rm K} r},
\end{align}
where $\Gamma = \Gamma_{\rm L} + \Gamma_{\rm c}$ is the total torque. We adopt
the prescription of \citet{Paardekooper_2011}, normalised to

\begin{equation}
\Gamma_0 = \left(\frac{q}{h}\right)^{2} \Sigma_{\rm gas}\, r^{4} \Omega_{\rm K}^{2},
\end{equation}
where $q = M_{\rm p}/M_\star$ and $h = H/r$. Writing the local slopes in the
standard positive convention, $\alpha = -\,{\rm d}\ln \Sigma_{\rm gas}/{\rm d}\ln r$
and $\beta = -\,{\rm d}\ln T/{\rm d}\ln r$, the Lindblad torque is given by

\begin{equation}
\gamma_{\rm eff}\,\frac{\Gamma_{\rm L}}{\Gamma_0} = -2.5 - 1.7\beta + 0.1\alpha.
\end{equation}
The corotation torque is the sum of barotropic and entropy-related horseshoe and linear contributions \citet{Paardekooper_2011}. Saturation is governed by the
viscous and thermal diffusion parameters, evaluated with
$\nu = \alpha_{\rm turb} c_{\rm s} H$ and the radiative diffusivity
\begin{equation}
\chi = \frac{16\,\gamma_{\rm ad}(\gamma_{\rm ad} - 1)\, \sigma_{\rm SB} T^{4}}
            {3\, \kappa\, \rho^{2} H^{2} \Omega_{\rm K}^{2}} ,
\end{equation}
where $\gamma_{\rm ad} = 1.4$ is the adiabatic index and $\rho$ the midplane gas density \citep{Bitsch_2011}. The opacity $\kappa$ entering $\chi$ is the Rosseland mean opacity of \citet{Bell_1994}, which coincides with the dust opacity used for the disc midplane temperature (Appendix~\ref{appendix:disc_model}).
The effective adiabatic index $\gamma_{\rm eff}$ interpolates between the
isothermal and adiabatic regimes following Eq.~(46) of
\citet{Paardekooper_2011}. Orbits are assumed circular and coplanar, so the
eccentricity and inclination damping factors are unity. Planets that migrate to radial distances smaller than $0.1$~AU are held at that radius, which we take as the inner edge of the
disc.

The balance between Lindblad and corotation torques determines both the direction and magnitude of migration. In particular, radial gradients in surface density and temperature can lead to regions of outward migration or migration traps.

\section{Model assumptions and caveats}
\label{appendix:assumptions}

The results presented in this study are subject to a number of simplifying assumptions 
that are worth making explicit. While the core-accretion framework adopted here 
captures the dominant physical processes governing gas giant formation, the fixed 
parameter choices inevitably reduce the variability inherent to real protoplanetary 
systems. The formation timescales shown in Fig.~\ref{fig.colourmap} should therefore 
not be interpreted as universal limits, but rather as indicative boundaries within a 
broader and more degenerate parameter space.

\begin{itemize}

    \item \textit{Grain opacity.} A reduction factor of $f_\mathrm{g} = 0.01$ is 
    adopted throughout. This low opacity significantly facilitates envelope contraction, 
    enabling crossover to be reached in the pebble-limited regime without hitting 
    isolation mass. Higher opacities would suppress this second formation pathway and 
    extend Kelvin-Helmholtz cooling timescales considerably.
\medskip
    \item \textit{Turbulent viscosity.} A fixed $\alpha = 10^{-4}$ is used. Lower values like $\alpha = 10^{-5}$ would accelerate accretion and likely 
    underestimate formation timescales. The not-reaching-isolation regime persists 
    across this range, but its extent is sensitive to the adopted value.
\medskip
    \item \textit{Disc mass and metallicity.} A disc mass of $10\%$ of the stellar mass 
    and a metallicity of $Z = 0.02$ are assumed. Higher disc masses and metallicities 
    would accelerate core growth and increase gas availability, systematically reducing 
    formation timescales. Moreover, a possible greater availability of solids at large separations might suppress the second pathway discussed in this study. Inversely, lower disc metallicity can show the second regime at closer separations.
\medskip
    \item \textit{Metallicity evolution.} The disc metallicity $Z = 0.02$ is kept constant for the entire disc lifetime, so we neglect its depletion by pebble drift and accretion. The pebble flux is computed from the initial gas surface density, $\Sigma_{\rm gas}(r, t=0)$, at the location of the outward-moving pebble-production front (Appendix~\ref{sec:accretion}), so it declines only as this front sweeps into the outer, less dense disc. This overestimates the pebble supply available at late times and large orbital distances relative to a model in which the local solid reservoir is genuinely exhausted. \citet{Gurrutxaga_2024} quantified this bias directly: benchmarking against the numerically self-consistent pebble flux of \citet{Appelgren_2023}, they showed that a constant-metallicity assumption systematically overestimates outer-disc core growth compared to a model with a self-consistently decaying flux, with the discrepancy increasing with disc age. Different values of the disc's metallicity would also affect the core growth and whether planetary cores can reach the isolation mass.
\medskip
    \item \textit{Migration.} Type~I migration is included, but the model does not 
    account for the trapping of pebbles at pressure bumps induced by an inward-drifting 
    planet, which would locally enhance pebble accretion and potentially alter the 
    crossover conditions.
\medskip
    \item \textit{Planetesimal enrichment.} No planetesimal accretion onto the core is 
    included. This could increase core masses, potentially pushing borderline cases 
    across the isolation mass threshold and modifying the core mass distribution 
    discussed in Section~\ref{Outer_disc}.
\medskip
    \item \textit{Solid composition.} Solids are assumed to be ice-dominated. 
    \citet{Valletta_2020} showed that water ice slightly overestimates formation 
    timescales relative to rock, which if anything strengthens the case for 
    sub-isolation crossover at large orbital distances presented here. Planet contraction and thus accretion is dependant on the EoS the model uses. 
\medskip
    \item \textit{Planet--planet interactions.} Only single-planet systems are 
    considered. Multi-body dynamics would introduce additional migration pathways and 
    make the orbital evolution considerably more degenerate.
\medskip
    \item \textit{Compositional homogeneity.} A distinct core--envelope structure is 
    assumed throughout, with accreted solids deposited at the core--envelope boundary. 
    In reality, ablation and fragmentation of infalling pebbles can distribute heavy 
    elements throughout the envelope, steepening the mean molecular weight gradient and 
    suppressing convection \citep{Brouwers_2021, MolLous_2024}. This slows 
    Kelvin--Helmholtz contraction relative to the sharp core--envelope case, extending 
    cooling timescales and pushing the viable formation window inward. The effect is most consequential in the pebble-exhaustion regime of Section~\ref{Outer_disc}, 
    where envelope contraction efficiency is already the limiting factor for reaching 
    crossover.
\medskip
    \item \textit{Embryo growth channel.} The initial growth to $0.1\,\mathrm{M}_\oplus$ is computed assuming solely pebble accretion. In reality, collisions between planetesimals could accelerate this stage. Therefore, the formation timescales presented in Fig.~\ref{fig.initial_timescale} should be taken as  upper bounds and the viable formation windows could be  somewhat wider.
\end{itemize}

Taken together, these assumptions define a controlled but idealised framework. The parameter choices are broadly representative of median disc conditions, and the qualitative trends, the stellar mass dependence of formation efficiency, the existence of two distinct accretion regimes, and the resulting core mass diversity are expected to be robust. Quantitative thresholds, however, will shift as more 
realistic disc physics, opacity models, and multi-planet architectures are incorporated. This naturally motivates a statistical population synthesis approach as a logical next step, where the parameter space explored here can be sampled more fully and confronted 
directly with the observed exoplanet demographics.

\end{appendix}
\end{document}